\documentclass[arxiv,usenatbib]{iupartex}
\usepackage{newtxtext,newtxmath}
\usepackage[T1]{fontenc}
\usepackage{ae,aecompl}

\usepackage{multicol}   % Multi-column entries in tables
\usepackage{bm}		    % Bold maths symbols, including upright Greek
\usepackage{pdflscape}	% Landscape pages

\title[NGC 188]{SED Analysis of the Old Open Cluster NGC 188}

\author[D. C. Dursun et al.]{%
D. C. Dursun$^{1\cc}$\orcid{0000-0001-7940-3731},
S. Ta\c {s}demir$^{1}$ \orcid{0000-0003-1339-9148}, 
S. Ko\c{ç}$^{1}$ \orcid{0000-0001-7420-0994}, and
S. Iyer $^{2}$\orcid{0009-0001-5220-0034}
\affsep \\
$^1$Istanbul University, Institute of Graduate Studies in Science, Programme of Astronomy and Space Sciences, 34116, Beyaz{\i}t, Istanbul, Turkey \\
$^2$ International Centre for Theoretical Sciences, 560089, Bengaluru, Karnataka, India
}
\corres{D. C. Dursun}{denizcdursun@gmail.com}

\pubyear{2024}

\doiheader{XXXXXXX/PAR.20XX.00000}
\date{
	\pSubmit{15.03.2024} 
	\pRevReq{08.04.2024}
	\pLastRevRec{16.04.2024}
	\pAccept{18.04.2024}
	\pPubOnl{XX.XX.XXXX}
}
\volume{0}
\volnumber{3}
\begin{document}
\label{firstpage}
\pagerange{\pageref*{firstpage}--\pageref*{lastpage}}
\maketitle
\begin{abstract}
In this study, we investigate the fundamental astrophysical parameters of the old open cluster NGC 188 through two complementary methods: isochron-fitting and spectral energy distribution (SED) analysis. Using photometric, astrometric, and spectroscopic data from the {\it Gaia} Data Release 3 \citep[DR3,][]{Gaia_DR3}, we identify 868 most likely member stars with membership probabilities $P \geq 0.5$. The mean proper-motion components and trigonometric parallaxes of the cluster are derived as ($\mu_{\alpha}\cos \delta$, $\mu_{\delta}$) = (-$2.314 \pm 0.002$, -$1.022 \pm 0.002$) mas yr$^{-1}$ and $\varpi = 0.550 \pm 0.023$, respectively. From this initial selection of high probable member stars, we proceed with the determination of astrophysical parameters using the isochron-fitting method. Simultaneously estimating the colour excess, distance, and age of the cluster, we employee PARSEC isochrones to observational data on {\it Gaia} based colour-magnitude diagrams. These findings were obtained as $E(G_{\rm BP}-G_{\rm RP})=0.066\pm 0.012$ mag, $d=1806 \pm21$ pc, and $t=7.65 \pm 1.00$ Gyr, respectively. Additionally, we identify and detected 19 previously confirmed blue straggler stars within NGC 188. Subsequently, we performed SED analyses for 412 out of the 868 cluster members. We obtained colour excess, distance and age of the cluster as $E(B-V)=0.034\pm 0.030$ mag, $d=1854\pm 148$ pc, and $t=7.78\pm 0.23$ Gyr, respectively. The analysis of member stars was revealed patterns of extinction in the $V$-band, with higher values of $A_{\rm V}$ observed in the lower right quadrant of the cluster. By comparing our results of SED analysis with models of stellar evolution, particularly in terms of temperature and surface gravity, we confirm agreement with theoretical predictions. This comprehensive investigation sheds light on the astrophysical properties of NGC 188, contributing to our understanding of stellar evolution within open clusters.

\end{abstract}


%%%%%%%%%%%%%%%%%%%%%%%%%%%%%%%%%%%%%%%%%%%%%%%%%%

%%%%%%%%%%%%%%%%% BODY OF PAPER %%%%%%%%%%%%%%%%%%

\section{Introduction}\label{sec:Introduction}

Open clusters (OCs) serve as invaluable natural laboratories to probe the fundamental principles of stellar evolution, Galactic dynamics, and the broader astrophysical processes that shape our Universe. OCs are groups of stars formed from the same molecular cloud under similar physical conditions.  Being gravitationally bound systems, OCs feature member stars that share similarities in terms of distance to the Sun, age, chemical composition, position, and velocity \citep{Harris_1994, Friel_1995, Lada_2003, Carraro_2007, Cantat-Gaudin-Anders_2020}. 

Recent advances in observational techniques have allowed scientists to delve deeper into the complexities of stellar phenomena, with spectral energy distribution (SED) analysis serving as a powerful tool for uncovering the fundamental properties of stars \citep{Lu_1999}. In astrophysics, SED analysis is a cornerstone method that provides deep insights into the fundamental properties of celestial objects, particularly stars. This analysis manifests in two primary forms: model-based and model-independent approaches. Serving as a link between theoretical frameworks and observational data, model-based SED analysis aids in determining critical parameters such as effective temperature, surface gravity, and metallicity for stars, thus enhancing our understanding of stellar evolution and behaviour. The examination of the SED of member stars in OCs yields essential astrophysical parameters of the cluster, which, in turn, can be utilized to decipher its dynamics and evolutionary trajectory \citep{Demarque_1992, Carraro_1994, VandenBerg_2004}. 

Each stellar constituent within OCs, categorized into different luminosity classes, undergoes a unique evolutionary trajectory shaped by factors such as mass, age, and chemical composition. Investigating the physical parameters across diverse evolutionary phases, from giants to main-sequence stars, allows for a comprehensive analysis of the intricate interplay of physical processes governing stellar evolution. The analysis of SED assumes paramount importance in OC studies, offering a means to scrutinize fundamental parameters of member stars originating from the same molecular cloud. Employing SED analysis on the most probable OC members enables the determination of crucial stellar characteristics, including effective temperature ($T_{\rm eff}$), surface gravity (log $g$), metallicity ([Fe/H]), $V$-band extinction ($A_{\rm V}$), distance ($d$), mass ($M$), radius ($R$), and age ($t$). Acquiring these fundamental physical parameters, particularly mass and radius, is essential for advancing our comprehension of stellar evolution. These analyses facilitate an in-depth exploration of the reddening effect on cluster members, shedding light on the influence of interstellar dust and gas.

With the beginning of the {\it Gaia} era \citep{Gaia_DR1}, high precision of astrometric data has been made available to the researchers, enabling accurate analyses for identifying the most likely cluster member stars. By comparing the colour-magnitude diagrams (CMDs) and two-colour diagrams (TCDs) of OCs with theoretical evolutionary models, important parameters such as age, distance, chemical composition, and interstellar extinction along the line of sight to the cluster can be determined. One traditional method used for this purpose is main-sequence fitting. This technique is based on the assumption that OCs members share similar characteristics, such as age, distance, and chemical composition, due to their common origin \citep{Carraro_1998, Chen_2003, Joshi_2005, Piskunov_2006, Joshi_2016}.

NGC 188 (Melotte 2, MWSC 0074) is an old open cluster that is located in a relatively low contaminated region of the Milky Way, making it an ideal object for observational studies. NGC 188 is located at $\alpha = 00^{\rm h} 47^{\rm m} 20^{\rm s}.96$ and  $\delta = +85^{\circ} 15^{\rm '} 05^{\rm''}.27$ (J2000.0), corresponding to Galactic coordinates of  $l=122^{\circ}.8368$ and $b = +22^{\circ}.3730$ \citep{Hunt_2023}. Several research efforts have analysed the main features of this open cluster, as detailed in Table \ref{tab:literature}. NGC 188 has a wide range of parameters, including ages ($t$) from 2.63 to 12 Gyr \citep{Demarque_1964, Hunt_2023}, metallicities ([Fe/H]) from -0.08 to 0.60 dex \citep{Spinrad_1970, Hills_2015}, colour excess ($E$($B$-$V$)) from 0.025 to 0.50 mag \citep{Sandage_1962, Fornal_2007}, and distances ($d$) from 1445 to 2188 pc \citep{Patenaude_1978, Hills_2015}.

\begin{table*}[h!]% Table 1
	\setlength{\tabcolsep}{7pt}
	\renewcommand{\arraystretch}{1.15}
	\small
	\centering
	\caption{Basic parameters of the NGC 188 open cluster collected from the literature.}\label{tab:literature}
	\begin{tabular}{cccccccc}
		\hline
		\hline
		$E(B-V)$ & $d$ & [Fe/H] & $t$ &  $\langle\mu_{\alpha}\cos\delta\rangle$ &  $\langle\mu_{\delta}\rangle$ & $V_R$ & Ref \\
		(mag) & (pc)  & (dex) & (Gyr) & (mas yr$^{-1}$) & (mas yr$^{-1}$) & (km s$^{-1})$ &      \\
		\hline
	    $0.50$                    &  $1549$         &$-$              & $-$          & $-$              & $-$              & $-$             & (01) \\
            0.1$\pm$0.020             &  $1500$         &$-$              & $-$          & $-$              & $-$              & -$49$           & (02) \\
            $-$                       &   $-$           &$-$              & $12$         & $-$              & $-$              & $-$             & (03) \\
            $0.18$                    &   $-$           &$-$              & $5.5$        & $-$              & $-$              & $-$             & (04) \\
            $0.15$                    &   $-$           &$0.60$           & $-$          & $-$              & $-$              & $-$             & (05) \\
            $0.15$                    &   $-$           &$-$              & $-$          & -$3.98$          & -$0.65$          & -$49$           & (06) \\
            $0.09$                    &   $-$           &$-$              & $-$          & $-$              & $-$              & $-$             & (07) \\
            $0.09$                    &   $1445$        &$-$              & $8$          & $-$              & $-$              & $-$             & (08) \\
            $-$                       &   $1700$        &$0.00$           & $6$          & $-$              & $-$              & $-$             & (09) \\
            $0.08$                    &   $-$           &$0.00$           & $10$         & $-$              & $-$              & $-$             & (10) \\
            $0.08$                    &   $-$           &  $-$            & $6.03$       & $-$              & $-$              & $-$             & (11) \\
            $0.12$                    &   $1995$        & 0.02$\pm$0.110  & $6$          & $-$              & $-$              & $-$             & (12) \\
            $0.03$                    &   $-$           & -0.06$\pm$0.00  & $7.5$        & $-$              & $-$              & $-$             & (13) \\
            $0.08$                    &   $1520$        & -$0.05$         & $7.2$        & $-$              & $-$              & $-$             & (14) \\
            0.09$\pm$0.020            &   $1905$        & -0.04$\pm$0.050 & 7$\pm$0.50   & $-$              & $-$              & $-$             & (15) \\
            $-$                       &  $-$            & 0.075$\pm$0.050 & $-$          & $-$              & $-$              & $-$             & (16) \\
            0.09$\pm$0.020            &  $-$            & $0.00$          & $6.8$        & $-$              & $-$              & $-$             & (17) \\
            0.025$\pm$0.005           &  1700$\pm$100   & $0.00$          & 7.5$\pm$0.70 & $-$              & $-$              & $-$             & (18) \\
            0.036$\pm$0.010           &  1714$\pm$64    & $0.12$          & 7.5$\pm$0.50 & $-$              & $-$              & $-$             & (19) \\
            $-$                       &  2188$\pm$100   & -0.08$\pm$0.003 & 6.45$\pm$0.04& $-$              & $-$              & $-$             & (20) \\
            0.033$\pm$0.030           &  1721$\pm$41    & $0.00$          & 7.08$\pm$0.04& $-$              & $-$              & -42.87$\pm$0.30 & (21) \\
            $-$                       &  $-$            & $0.00$          & $6$          & $-$              & $-$              & $-$             & (22) \\
            $0.075\pm0.008$           &  $-$            & $-$             & $-$          &~-3.000$\pm$1.830 & -0.370$\pm$0.100 & $-$             & (23) \\
            $-$                       &  $1864\pm$4     & $-$             & $-$          &-2.307$\pm$0.139  & -0.960$\pm$0.146 & $-$             & (24) \\
            $-$                       &  $1864\pm$4     & $-$             & $-$          &-2.307$\pm$0.139  & -0.960$\pm$0.146 & -41.70$\pm$0.19 & (25) \\
            $-$                       &  $-$            & 0.14$\pm$0.003  & $4.47$       &-2.310$\pm$0.190  & -0.960$\pm$0.160 & -41.50$\pm$1.10 & (26) \\
            0.068                     &  1698           & $-$             &$7.08$        &-2.307$\pm$0.139  & -0.960$\pm$0.146 &  $-$            & (27) \\
            $-$                       &  1859$\pm$36    &0.112$\pm$0.020  &$7.05$        &-2.302$\pm$0.184  & -0.955$\pm$0.172 &-41.602$\pm$0.58 & (28) \\
            $-$                       &  1974$\pm$20    &0.088$\pm$0.032  &$7.08$        &-2.303$\pm$0.182  & -0.953$\pm$0.167 & $-$             & (29) \\
            $-$                       &  1698           &0.088$\pm$0.032  &$7.08$        & $-$              & $-$              &-42.03$\pm$0.05  & (30) \\
            $-$                       &  $-$            &$-$              &$-$           & $-$              & $-$              &-41.64$\pm$0.25  & (31) \\ 
            $-$                       &  1670           &0.090$\pm$0.020  &$7.59$        & $-$              & $-$              & $-$             & (32) \\
            $-$                       &  1847$\pm$6     &$-$              &$-$           & -2.335$\pm$0.004 & -1.024$\pm$0.004 & -41.70$\pm$0.20   & (33) \\
            0.074$\pm$0.037           &  $1822$         & $-$             &2.63$\pm$1.17 &-2.318$\pm$0.106  & -1.015$\pm$0.111 &-41.13$\pm$0.59  & (34) \\
            0.047$\pm$0.009           &  1806$\pm$21    & $-$             &7.65$\pm$1.00 &-2.314$\pm$0.048  & -1.020$\pm$0.045 &-41.59$\pm$0.14  & (35) \\
		\hline  
	\end{tabular}%
	\\

\begin{minipage}{15cm}
(01)~\citet{Sandage_1962}, (02)~\citet{Greenstein_1964}, (03)~\citet{Demarque_1964}, (04)~\citet{Aizenman_1969}, (05)~\citet{Spinrad_1970}, (06)~\citet{Upgren_1972}, (07)~\citet{McClure_1977}, (08)~\citet{Patenaude_1978}, (09)~\citet{Vandenberg_1983}, (10)~\citet{Vandenberg_1983}, (11)~\citet{Janes_1983}, (12)~\citet{Caputo_1990},  (13)~\citet{Carraro_1994}, (14)~\citet{Friel_1995}, (15)~\citet{Sarajedini_1999}, (16)~\citet{Worthey_2003}, (17)~\citet{VandenBerg_2004}, (18)~\citet{Fornal_2007}, (19)~\citet{Wang_2015}, (20)~\citet{Hills_2015}, (21)~\citet{Elsanhoury_2016}, (22)~\citet{Lorenzo-Oliveira_2016}, (23)~\citet{Dias_2014}, (24)~\citet{Cantat-Gaudin_2018}, (25)~\citet{Soubiran_2018}, (26)~\citet{Donor_2018}, (27)~\citet{Cantat-Gaudin_2020}, (28)~\citet{Dias_2021}, (29)~\citet{Spina_2021}, (30)~\citet{Tarricq_2021}, (31)~\citet{Carrera_2022}, (32)~\citet{Netopil_2022}, (33)~\citet{Gao_2022}, (34)~\citet{Hunt_2023}, (35)~This study
\end{minipage}
\end{table*}%

\citet{Bruce_1989} compared the ages of NGC 188 and M67 and found that NGC 188 is slightly older OC, at approximately 6.5 Gyr. They suggested adjustments for reddening and distance modulus to reconcile inconsistencies and remove anomalies, such as the lithium discrepancy. \citet{Leonard_1992} explored the origins of blue stragglers and contact binaries in M67 and NGC 188, proposing physical stellar collisions and tidal captures as potential mechanisms. The experiments showed that these interactions could explain approximately 10$\%$ of the observed objects. \citet{Belloni_1998} conducted X-ray observations of M67 and NGC 188, detecting various sources and noting puzzling emissions from specific binaries in M67. They also identified two members in NGC 188, including the FK Com type star D719. \citet{Glebbek_2008} investigated the detailed evolution of stellar collision products in OCs, with a particular focus on M67 and NGC 188. The authors presented models of merger remnants and compared them to observed blue straggler populations, indicating recent collision events in M67. In a photometric survey of NGC 188, \citet{Song_2023} identified 25 variable stars, including one new variable star, and discussed their characteristics, such as spectral types and classifications, providing insights into the cluster's stellar population.

This paper aims to determine the fundamental parameters that define the old open cluster of NGC 188 using advanced analytical techniques, such as isochrone fitting and SED analyses. It is important to acknowledge the potential for parameter degeneracy in analysis processes, which can complicate the investigation. To deal with this challenge, we adopt a rigorous approach by calibrating the distances with trigonometric parallaxes from the {\it Gaia} Data Release 3 \citep[{\it Gaia} DR3, ][]{Gaia_DR3} catalogue and taking advantage of metallicity values from high-resolution spectroscopic data in the literature.  These steps aim to minimise parameter degeneracy and provide an accurate determination of the age and other fundamental parameters of the NGC 188 open cluster.

The subsequent sections of this paper are structured as follows: A description of the astrometric and photometric data of NGC 188 is given in Section \ref{sec:Data}. In Section \ref{sec:Method}, the presented methods are used to derive the fundamental parameters of NGC 188. In Section \ref{sec:Results}, the photometric membership probabilities and structural parameters of the stars in NGC 188 are presented and discussed, followed by the main astrophysical parameters obtained with SED. Finally, Section \ref{sec:Conclusion} provides a summary.

\section{Data}\label{sec:Data}

\subsection{Photometric and Astrometric Data}
\begin{figure}[t!]%Figure 01
	\centering \includegraphics[width=0.9\linewidth]{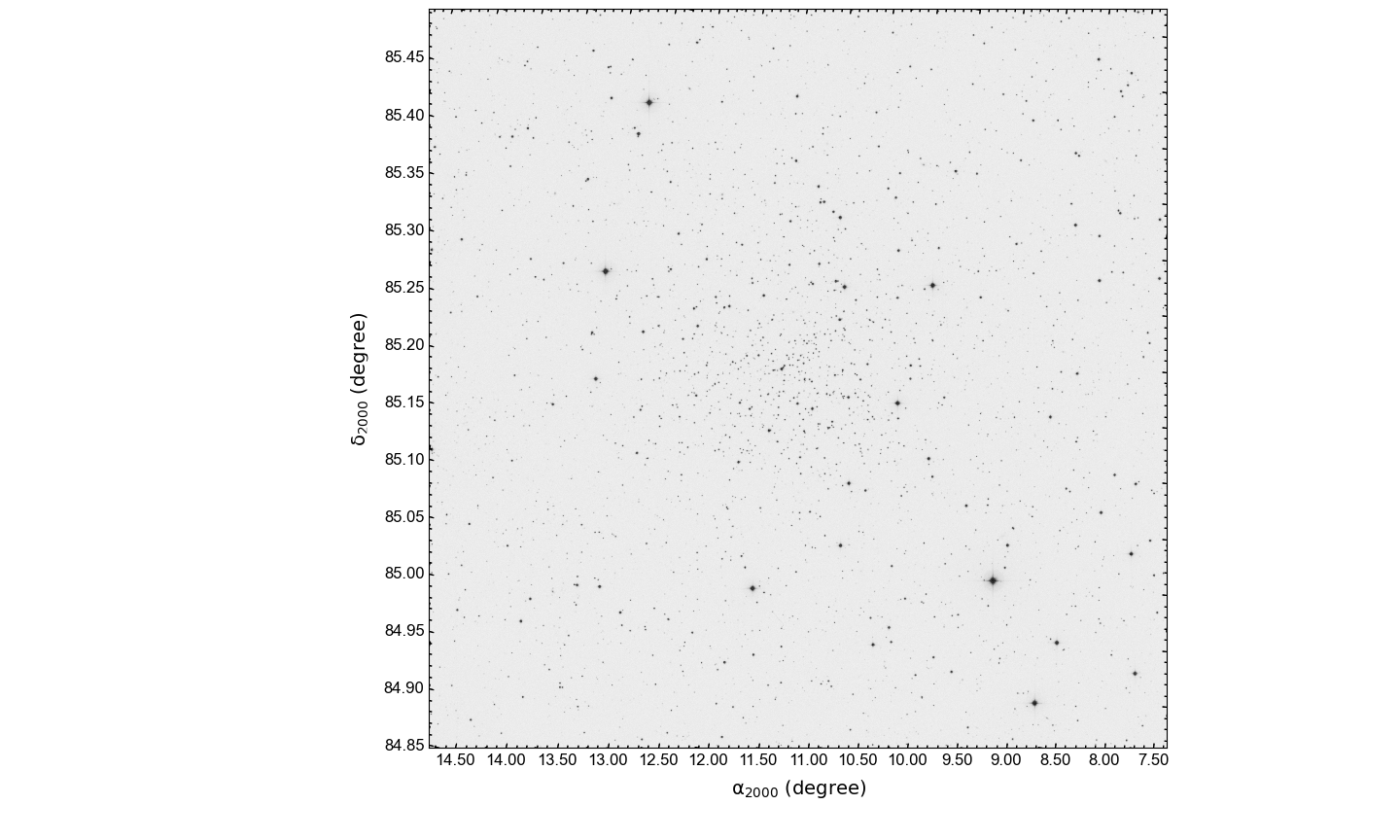}\vspace*{-1.5mm}
	\caption{ Identification chart of the NGC 188 for $40'\times 40'$ region. Up and left directions represent the North and East, respectively.} 
	\label{fig:ID_charts}
\end{figure}

The photometric and astrometric analyses of NGC 188 utilized data from the {\it Gaia} DR3 catalogue \citep[DR3,][]{Gaia_DR3}. Astrometric and photometric data were generated based on the equatorial coordinates provided by \citet{Hunt_2023} ($\langle\alpha, \delta\rangle) = (00^{\rm h} 47^{\rm m} 20^{\rm s}.96, \delta= +85^{\circ} 15^{\rm '} 05^{\rm''}.27$). Encompassing the entire field of NGC 188, all stars within a $40\arcmin$ radius from the cluster center were considered. Consequently, 17,344 stars falling within the $6<G (\rm mag) \leq23$ mag range were detected. The identification chart of stars in the direction of NGC 188, covering a $40\arcmin \times 40 \arcmin$ field, is illustrated in Figure~\ref{fig:ID_charts}.

\begin{figure}[t!]%Figure 02
	\centering \includegraphics[width=0.9\linewidth]{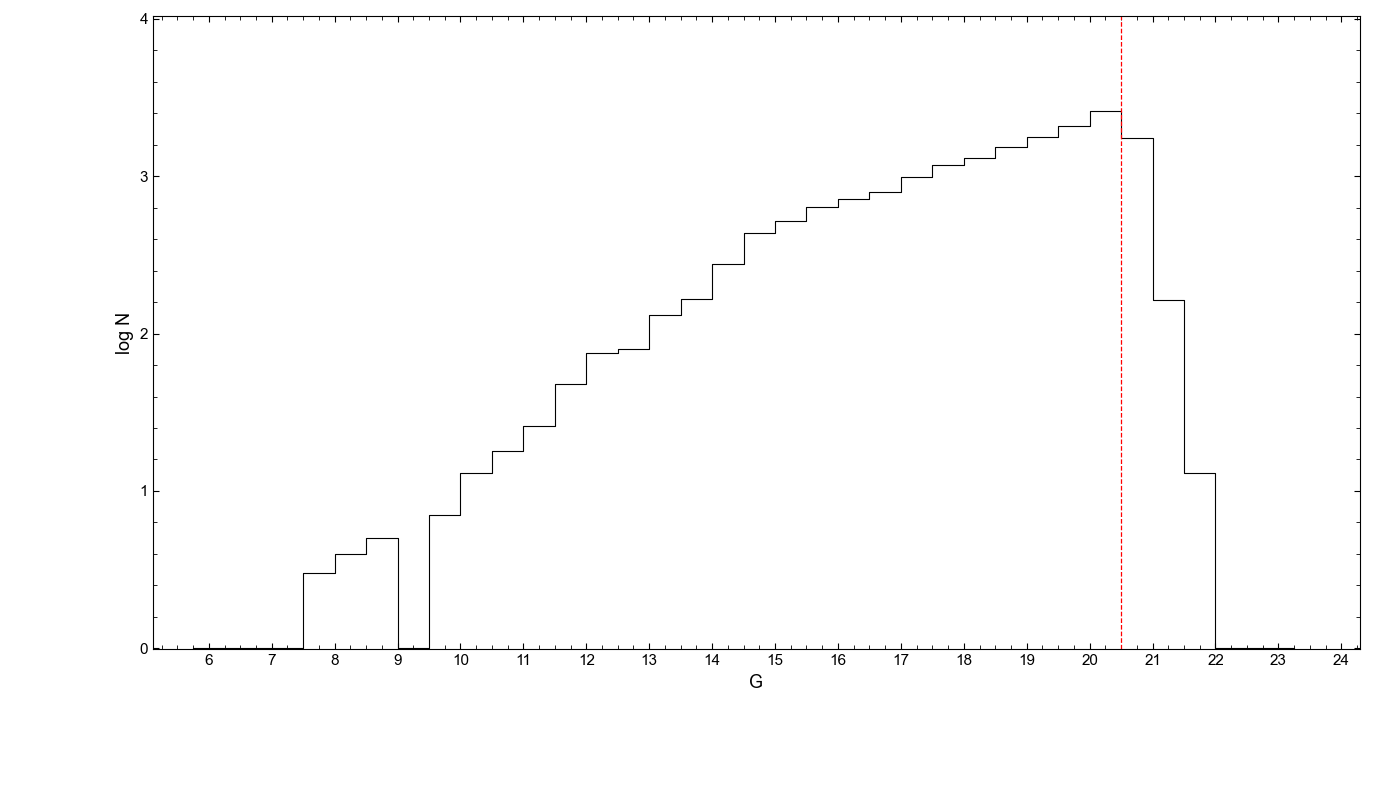}\vspace*{-1.5mm}
	\caption{Distribution of the stars in the direction of NGC 188 for $G$ magnitude intervals. The photometric completeness limit is indicated by a red dashed line.}
	\label{fig:histograms1}
\end{figure}

\subsection{Photometric Completeness Limit and Errors}

To precisely determine the structural and astrophysical parameters of the cluster, it is imperative to establish the photometric completeness limit by tallying stars corresponding to $G$ magnitudes. The photometric completeness limit is defined as the magnitude where the number of stars increases with magnitude up to a certain point, beyond which it starts to decrease. For NGC 188, this value, as evident from the histogram in Figure~\ref{fig:histograms1}, is the photometric completeness limit of $G=20.5$ mag. Stars fainter than this completeness limit were excluded, ensuring they were not considered in subsequent analyses. Photometric inaccuracies reported in {\it Gaia} DR3 were treated as internal errors, reflecting uncertainties associated with the instrumental magnitudes of celestial bodies. Consequently, the study considered uncertainties of the instrumental magnitudes of stars as internal errors. Mean errors for $G$ magnitudes and $G_{\rm BP}-G_{\rm RP}$ colour indices were computed across the $G$ apparent magnitude interval. The mean errors for $G$ magnitudes along with $G_{\rm BP}-G_{\rm RP}$ colour indices of stars are presented in Table \ref{tab:photometric_errors} as a function of $G$ magnitudes. At the mean internal error for $G$ magnitude and $G_{\rm BP}-G_{\rm RP}$ colour index were determined to be 0.009 and 0.182 mag, respectively.

\begin{table}%Table 02
\setlength{\tabcolsep}{10pt}
  \centering
  \caption{Mean internal photometric errors of NGC 188 for $G$ and $G_{\rm BP}-G_{\rm RP}$ mag in per $G$ magnitude bin.}
    \begin{tabular}{cccc}
      \hline
  $G$ & $N$ & $\sigma_{\rm G}$ & $\sigma_{G_{\rm BP}-G_{\rm RP}}$\\
(mag) &  & (mag) & (mag) \\
\hline
    (06, 14]	& ~~487 & 0.002   & 0.005 \\
    (14, 15]	& ~~593	&  0.002  &	0.005 \\
    (15, 16]	& 1035	&  0.002  &	0.006 \\
    (16, 17]	& 1447	&  0.002  &	0.008 \\
    (17, 18]	& 1969	&  0.003  &	0.017 \\
    (18, 19]	& 2598	&  0.003  &	0.040 \\
    (19, 20]	& 3573	&  0.004  &	0.086 \\
    (20, 21]	& 4988	&  0.009  &	0.182 \\
    (21, 23]	& ~~654	&  0.026  &	0.297 \\
   \hline
    \end{tabular}%
  \label{tab:photometric_errors}%
\end{table}%

\section{Method}\label{sec:Method}
\subsection{Isochrone Fitting}

A classical technique employed for determining the fundamental astrophysical parameters of open clusters is the main-sequence fitting method. This method relies on the assumption that members of a star cluster originate from the same molecular cloud and share common properties such as distance, age, and chemical composition. The isochrone fitting method involves comparison with theoretical isochrones to simultaneously determine the age, metallicity, isochrone distance of the cluster, and the best isochrone repsresenting the cluster is shown by fitting over the CMD.  The isochrone fitting process entails selecting isochrones with different ages and metallicities that best fit the observed CMD of cluster members. However, this process may introduce parameter degeneracy \citep{King_2005, deMeulenaer_2013, Janes_2014}.

To mitigate potential degeneracy in the analyses, we imposed constraints on the distance and metallicity parameters of the cluster. The distance was chosen proximate to the value calculated from the mean trigonometric parallax of NGC 188. Additionally, the metallicity of the cluster was derived from a literature study that provided high-resolution spectroscopic data. This approach was adopted to minimize parameter degeneracy in cluster analyses \citep[cf.][]{Yontan_2015, Yontan_2019, Yontan_2023}.

\subsection{Spectral Energy Distribution}

In astrophysics, spectral energy distribution (SED) analysis aims to determine the physical properties of stars and other astronomical objects by examining the wavelength and intensity distributions of the light emitted by them across the electromagnetic spectrum \citep{Oke_1974, Adams_1987, Thomas_2006, Yadav_2024}. SED requires observational fluxes with different filters to measure the object over a wide range of the electromagnetic spectrum. These observed fluxes are then analyzed to determine the astrophysical parameters of the object by comparison with theoretical models. SED is assisted by computer simulations and utilizes optimization techniques to effectively match the properties of the astronomical object while considering the complexities of the observed data. In particular, SED analysis determines astrophysical parameters ($T_{\rm eff}$, $\log g$, [Fe/H], $A_{\rm v}$, $d$, $M$, $R$, and $t$) of stars. This technique is used to understand various astrophysical topics, such as star formation and evolution, Galactic and cosmic evolution processes, and galaxy formation. Furthermore, SED analysis of member stars in open clusters is used to determine the age, chemical composition, and evolutionary state of the star clusters.

For the SED method, we utilized the SpectrAl eneRgy dIstribution bAyesian moDel averagiNg fittEr \citep[ARIADNE;][]{Vines_2022} code for member stars with photometric data points covering a wavelength range from UV to IR. {\sc ARIADNE} has been designed with a focus on speed, user-friendliness, and versatility. It employs a Bayesian framework to estimate physical characteristics and associated uncertainties efficiently. The platform is flexible and can accommodate various stellar evolution models, star formation scenarios, dust attenuation profiles, and the inclusion of nebular emissions. Furthermore, it provides a $\chi^2$ minimization feature through {\sc ARIADNE}, facilitating straightforward comparisons with existing research. {\sc ARIADNE} is particularly well-suited for investigating stellar clusters. More than 20 photometric data points within the wavelength range $0.1<\lambda <5~\mu {\rm m}$ of the electromagnetic spectrum are used to fit the SED of stars. {\sc ARIADNE} determines the astrophysical parameters of single stars. For the synthetic models included in {\sc ARIADNE}, the three models with the widest effective temperature, surface gravity, and metallicity parameter range ($2300<T_{\rm eff} (\rm K) <12000$, $0<\log g (\rm cgs)<6$, and -$2.5<{\rm [Fe/H] (dex)}< 1$) PHOENIX v2 \citep{Huser_2013}, BT-Cond \citep{Allard_2012}, Castelli and Kurucz \citep{Castelli_2003} stellar atmosphere models were used.

\section{Results}\label{sec:Results}
\subsection{Structural Parameters}

Radial Density Profile (RDP) analysis is utilized to determine the spatial extent of NGC 188 and obtain its structural parameters. The cluster area is divided into numerous concentric rings, considering the central coordinates provided by \citet{Hunt_2023} through {\it Gaia} DR3 data within a 40$\times$40 arcmin$^{2}$ region. To compute the stellar density ($\rho(r)$) of NGC 188, stars within the $G \leq 20.5$ mag completeness limit are considered, and the equation $R_{i}=N_{i}/A_{i}$ is applied for each $\it i^{\rm th}$ ring, where $N_{i}$ and $A_{i}$ denote the number of stars falling into a ring and the area of the particular ring, respectively. Then, calculated stellar densities were plotted against the distance from the centre of NGC 188 and an empirical King profile \citet{King_1962} fitted which is defined as $\rho (r)= f_{\rm bg}$ + $f_{\rm 0}$ / (1 + $(r/+ $$r_{\rm c}$$)^{2}$). Here, $r$ and $r_{\rm c}$ imply the angular radius and core radius, respectively, as well as the $f_{\rm bg}$ and $f_0$ represent the background stellar density and central stellar density, respectively.

\begin{figure}[b!]%Figure 03
	\centering
	\includegraphics[width=0.6\linewidth]{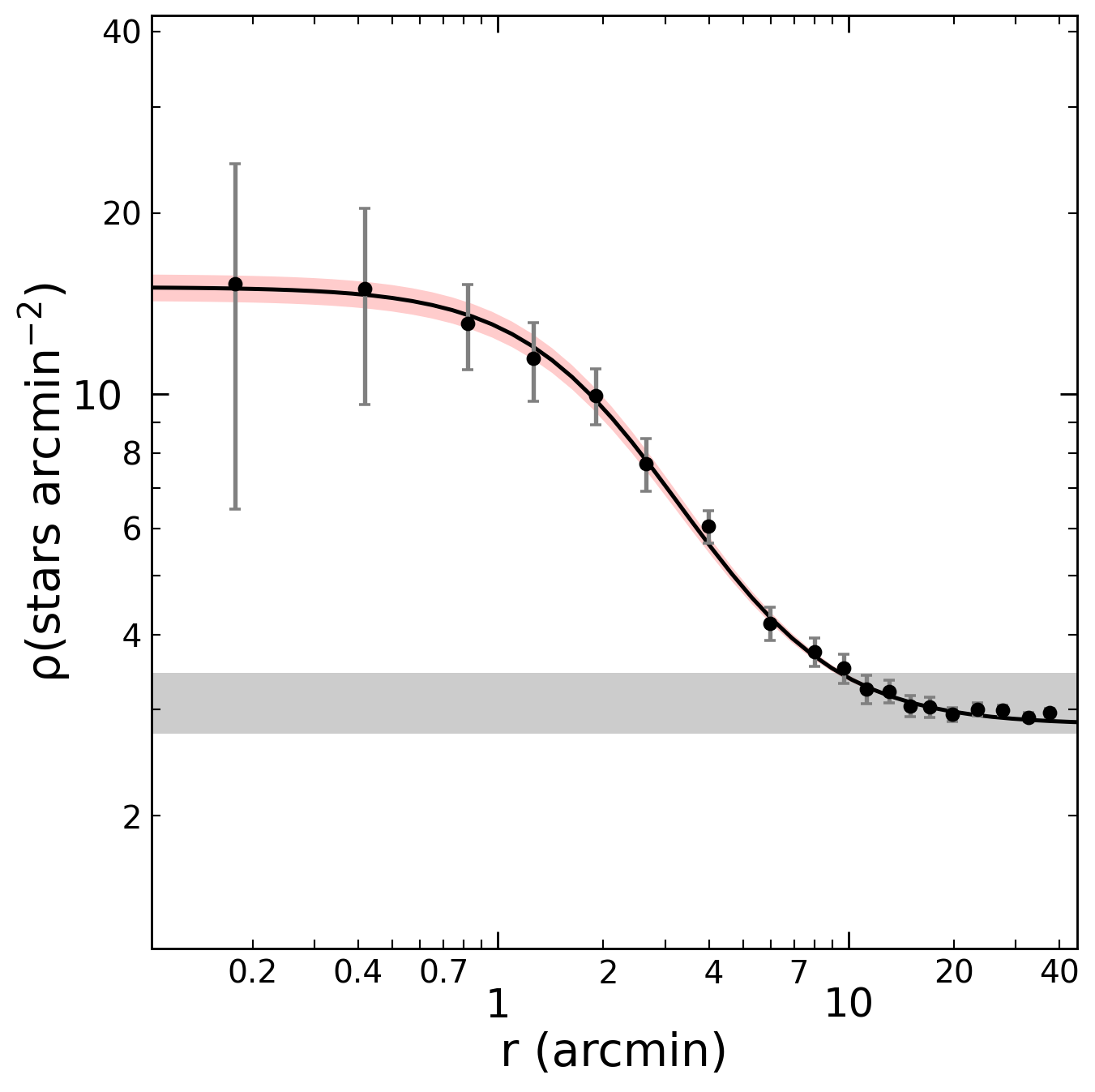}
	\caption{The RDP of \citet{King_1962} for NGC 188. Stellar density errors were determined from Poisson statistics $1/\sqrt N$, where $N$ is the number of stars. The fitted black curve and horizontal grey shaded area show the best-fitted RDP and background stellar density, respectively. Also, red-shaded area indicates the $1\sigma$ uncertainty of the King fit.} 
	\label{fig:king}
\end{figure}

The RDP fitting method employed the $\chi^{2}$ minimization technique, and the best-fit solution of the RDP is depicted with a black solid line in Figure \ref{fig:king}. Examining the figure reveals that the stellar density of NGC 188 peaks around the center of the cluster, gradually decreasing radially as it moves away from the cluster center. The RDP flattens and merges with the background star density at a specific point known as the limiting radius. In this study, we estimated this radius through visual inspection and adopted it as $15'$ (Figure \ref{fig:king}). The stars located within this observational limiting radius were utilized in further analyses. To confirm the reliability of the observed limiting radii ($r_{\rm lim}^{\rm obs}$) by theoretical approach, we used the equation given by \citet{Bukowiecki_2011} that is expressed by 
$r_{\rm lim}^{\rm teo}=r_{\rm c}$ $\times$ (($f_{\rm 0}$ / ${3\sigma_{\rm bg}})-1)^{1/2}$. Considering this equation, the theoretical limiting radius is calculated as $r_{\rm lim}^{\rm teo}=14\arcmin.8$. It is clear to see that theoretical and observed limiting radii values are in good agreement. The central and background stellar density, as well as the core radius of NGC 188 are obtained as $f_0=12.229\pm0.768$ stars arcmin$^{-2}$, $f_{\rm bg}=2.832\pm0.356$ stars arcmin$^{-2}$ and $r_{\rm c}=2\arcmin.183\pm0\arcmin.304$, respectively. Results are listed in Table \ref{tab:Final_table}.

\subsection{Membership Probabilities and Astrometric Analysis}\label{sec:Membership}

To accurately determine the astrophysical parameters of Open Clusters (OCs), it is crucial to distinguish the physical members of the cluster from the field stars, given the significant impact of field stars on OCs located in the Galactic plane. The members of an open cluster share the same origin, arising from the collapse of a common molecular cloud. Consequently, the proper-motion vectors of cluster member stars exhibit a consistent direction in space, and their proper motion values closely align with the mean proper motions of the cluster. This congruence serves as a valuable tool for effectively separating field stars from cluster stars.

\begin{figure}%Figure 04
	\centering
	\includegraphics[width=0.6\linewidth]{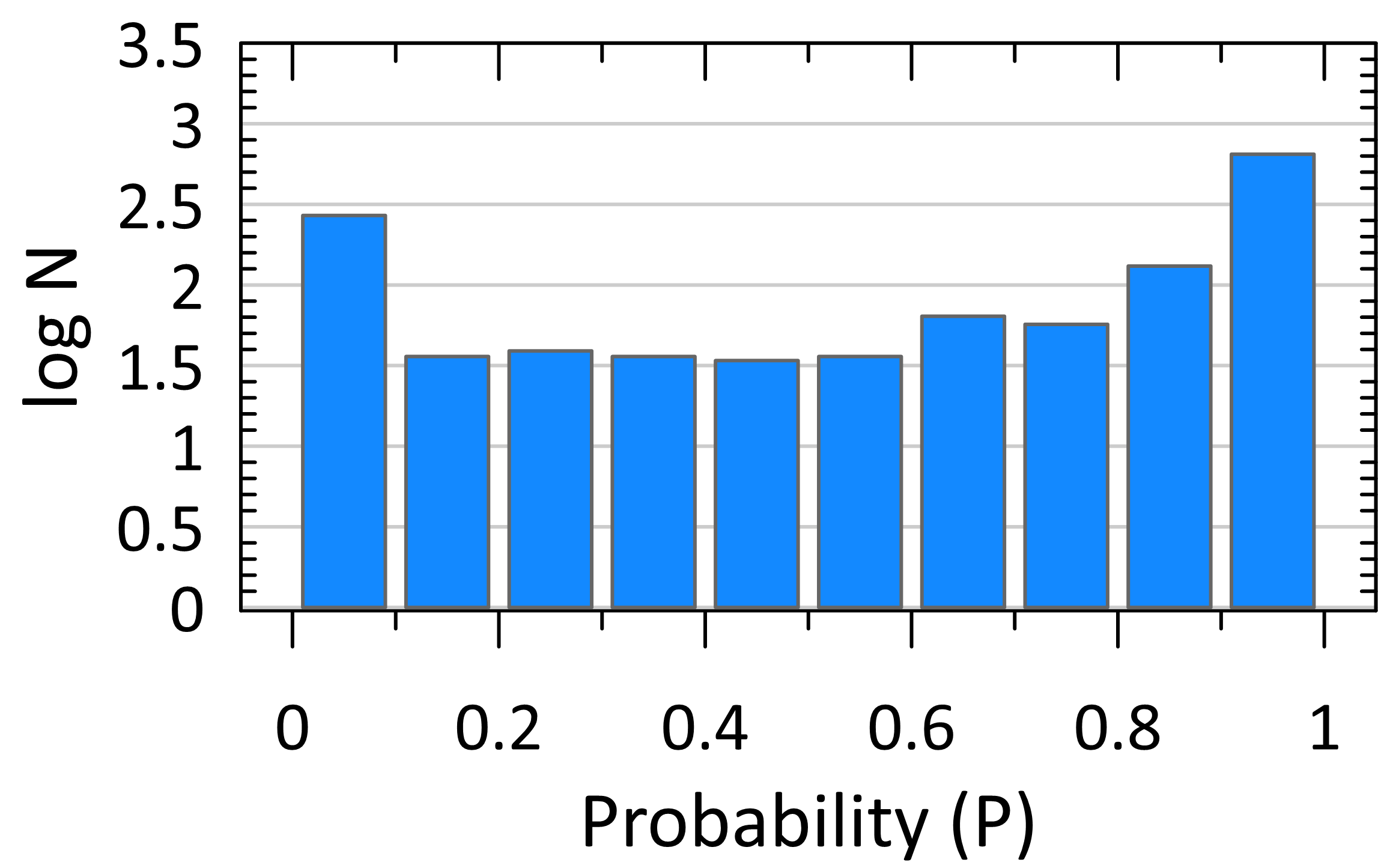}
	\caption{Distribution of cluster membership probabilities for the stars in the direction of NGC 188.} 
	\label{fig:NGC_188_Prob}
\end{figure}

The Unsupervised Photometric Membership Assignment in Stellar Clusters ({\sc upmask}) method was employed for membership analyses, utilizing astrometric parameters, including equatorial coordinates ($\alpha$, $\delta$), proper-motion components ($\mu_{\alpha}\cos \delta$, $\mu_{\delta}$), and trigonometric parallaxes ($\varpi$), along with their uncertainties, from the {\it Gaia} DR3 catalogue of NGC 188. {\sc upmask} relies on a machine-learning clustering algorithm, specifically $k$-means clustering, to identify similar groups of stars based on their proper motion components and trigonometric parallaxes. This approach facilitates the statistical determination of members within the open cluster. The membership probability histogram is shown in Figure \ref{fig:NGC_188_Prob}. Stars with membership probabilities $P\geq 0.5$ were considered potential cluster members. Through astrometric calculations and considering photometric limitations, 868 stars were identified as the most probable physical members for NGC 188. These stars not only have membership probabilities $P\geq 0.5$ but are also within the observational limiting radius ($r_{\rm lim}^{\rm obs}$) and satisfy the photometric completeness limit ($G\leq 20.5$ mag).
  
\begin{figure}[t!]%Figure 05
	\centering \includegraphics[width=1\linewidth]{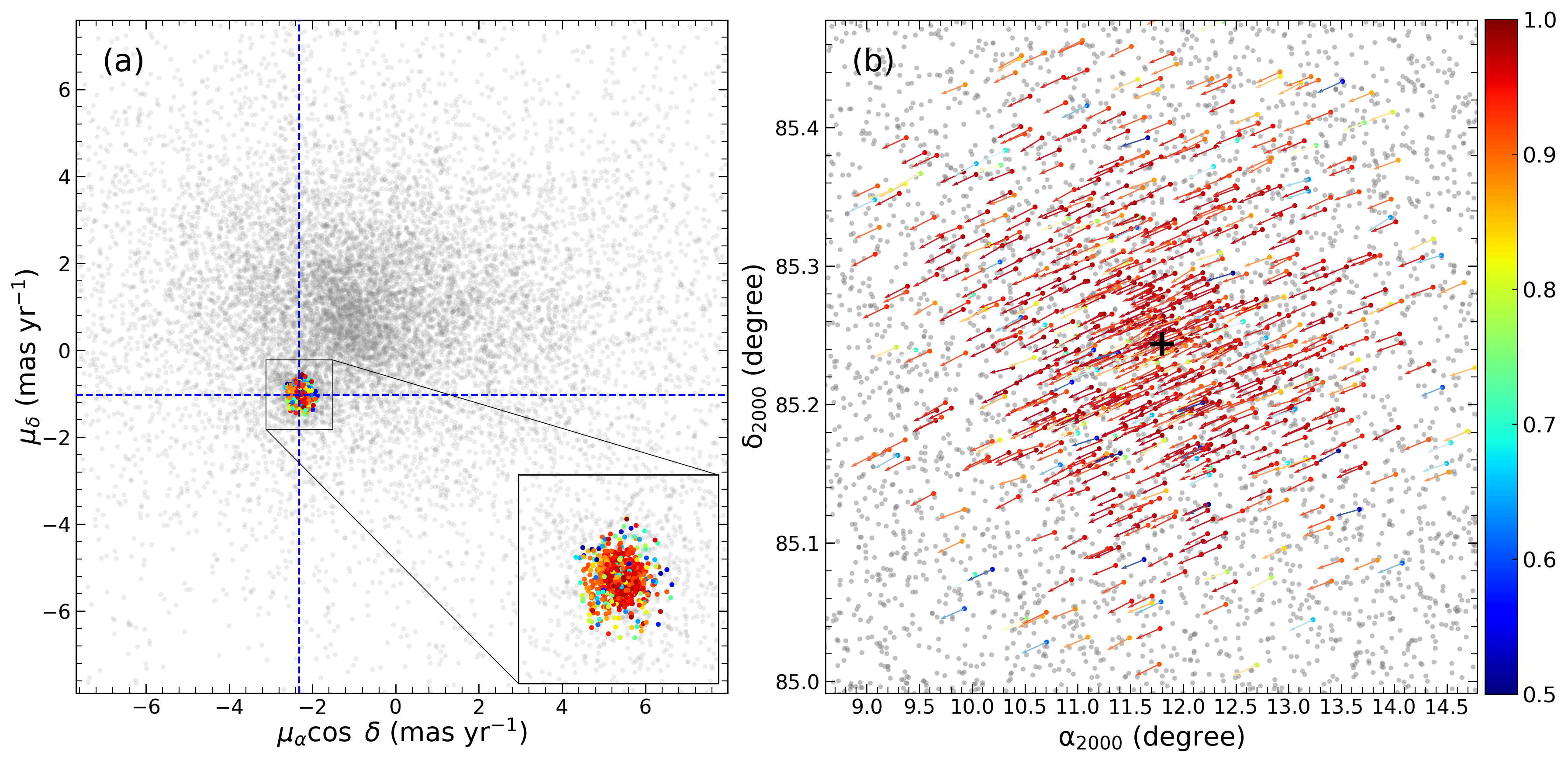}
	\caption{Vector Point Diagram (a) and proper-motion velocity vectors (b) of NGC 188. Colour scale in the right panel denotes the membership probabilities over than 0.5. In panel (a), the magnified boxes reveal regions with a high concentration of member stars in the VPDs, and mean proper-motion values are indicated by the intersection of blue dashed lines. The centre of equatorial coordinates of the NGC 188 are marked by black cross-hairs in panel (b).
		\label{fig:VPD_all}} 
\end{figure}

We computed the mean proper-motion components of the cluster for stars with membership probabilities $P\geq 0.5$ and illustrated their distribution throughout the cluster using the vector point diagram (VPD) in Figure \ref{fig:VPD_all}. The cluster occupies a distinct region, relatively separated from field stars. The mean values of the proper-motion components specific to the cluster are represented at the intersection of the blue dashed lines. The calculated mean proper-motion components for NGC 188 are ($\mu_{\alpha}\cos \delta$, $\mu_{\delta}$) = (-$2.314 \pm 0.002$, -$1.022 \pm 0.002$) mas yr$^{-1}$, aligning well with recent studies in the literature \citep[e.g.,][]{Cantat-Gaudin_2020, Dias_2021, Hunt_2023}. 

While trigonometric parallax measurements represent the most precise method for determining stellar distances, the existence of errors at the zero point in astrometric measurements introduces considerable uncertainty, particularly in distance determinations for distant objects. Recent studies \citep[e.g.,][]{Lindegren2021, Huang2021, Riess2021, Zinn2021} have proposed zero-point corrections utilizing a multitude of objects with trigonometric measurements in the {\it Gaia} EDR3/DR3 database \citep{Gaia_EDR3, Gaia_DR3}. Given that NGC 188 is situated at a distance of approximately 1.8 kpc \citep{Dias_2021, Hunt_2023}, we applied a zero-point correction to the trigonometric parallaxes ($\varpi$) of the most likely cluster members ($P \geq 0.5$). This correction involved considering the value $\varpi_{\rm ZP}=-0.025$ mas, as proposed by \citet{Lindegren2021}, and employing the relation $\varpi_0 = \varpi - \varpi_{\rm ZP}$ for each member star.

\begin{figure}[h!]%Figure 06
	\centering \includegraphics[width=0.5\linewidth]{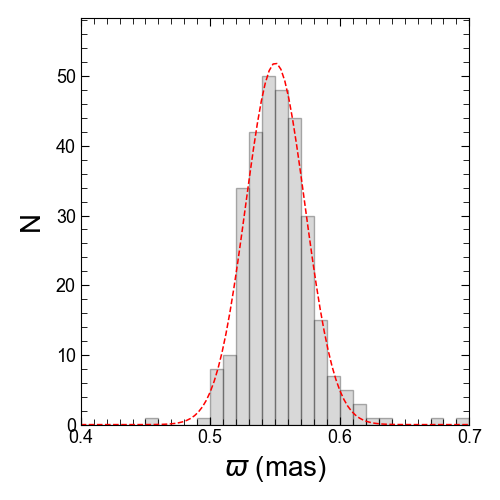}
	\caption{{\it Gaia} DR3-based trigonometric parallax histogram constructed from the most likely member stars of NGC 188. The Gaussian fit applied to the distributions is represented by red-dashed curve.
		\label{fig:plx_hist}}
\end{figure}

To calculate the mean trigonometric parallax of NGC 188, we focused on stars with a relative parallax error smaller than 0.05. A histogram of the trigonometric parallaxes for the most likely members was plotted, and a Gaussian function was fitted to determine the mean trigonometric parallax of the cluster, as illustrated in Figure \ref{fig:plx_hist}. The mean trigonometric parallax of NGC 188 was determined as $\varpi$ = 0.550 $\pm$ 0.023 mas. The linear distance of the cluster was computed using the equation $d({\rm pc})=1000/\varpi$. Consequently, the transformed trigonometric parallax yielded an estimated distance of $d_{\varpi}=1818\pm 76$ pc. We found this result to be in good agreement with values reported in the literature \citep{Sarajedini_1999, Cantat-Gaudin_2018}.

\subsection{Astrophsical Parameters}
\subsubsection{Isochrone Fitting Method}

In determining the fundamental astrophysical parameters such as reddening, age, and distance of OCs, CMDs can be used as an important tools. In this study, most likely stars were projected on the $G\times (G_{\rm BP}-G_{\rm RP})$ CMD with their membership probabilities. The mean metallicity of the NGC 188 was taken directly from literature to avoid parameter degeneration. We adopted the value of \citet{Casamiquela_2021} who determined the mean metallicity of the NGC 188 as -$0.030\pm0.015$ dex by analysing the high-resolution spectra of four member stars. In order to select the best-fit isochrone and obtain the astrophysical parameters, adopted metallicity ([Fe/H]=-0.030 $\pm$ 0.015 dex) is converted to mass fraction $z$ by using the equation given by Bovy\footnote{https://github.com/jobovy/isodist/blob/master/isodist/Isochrone.py} that are available for {\sc PARSEC} models \citep{Bressan_2012}. 

\begin{equation}
z_{\rm x}={10^{{\rm [Fe/H]}+\log \left(\frac{z_{\odot}}{1-0.248-2.78\times z_{\odot}}\right)}}
\end{equation}
and
\begin{equation}
z=\frac{(z_{\rm x}-0.2485\times z_{\rm x})}{(2.78\times z_{\rm x}+1)}.
\end{equation} 
where $z_{\rm x}$ and $z_{\odot}$ are intermediate values where solar metallicity $z_{\odot}$ was adopted as 0.0152 \citep{Bressan_2012}. Using these equations we derived the  mass fraction value that corresponds to [Fe/H] = -$0.030\pm0.015$ dex  as $z=0.0142$.

\begin{figure}[t!]%Figure 07
\centering\includegraphics[width=0.5\linewidth]{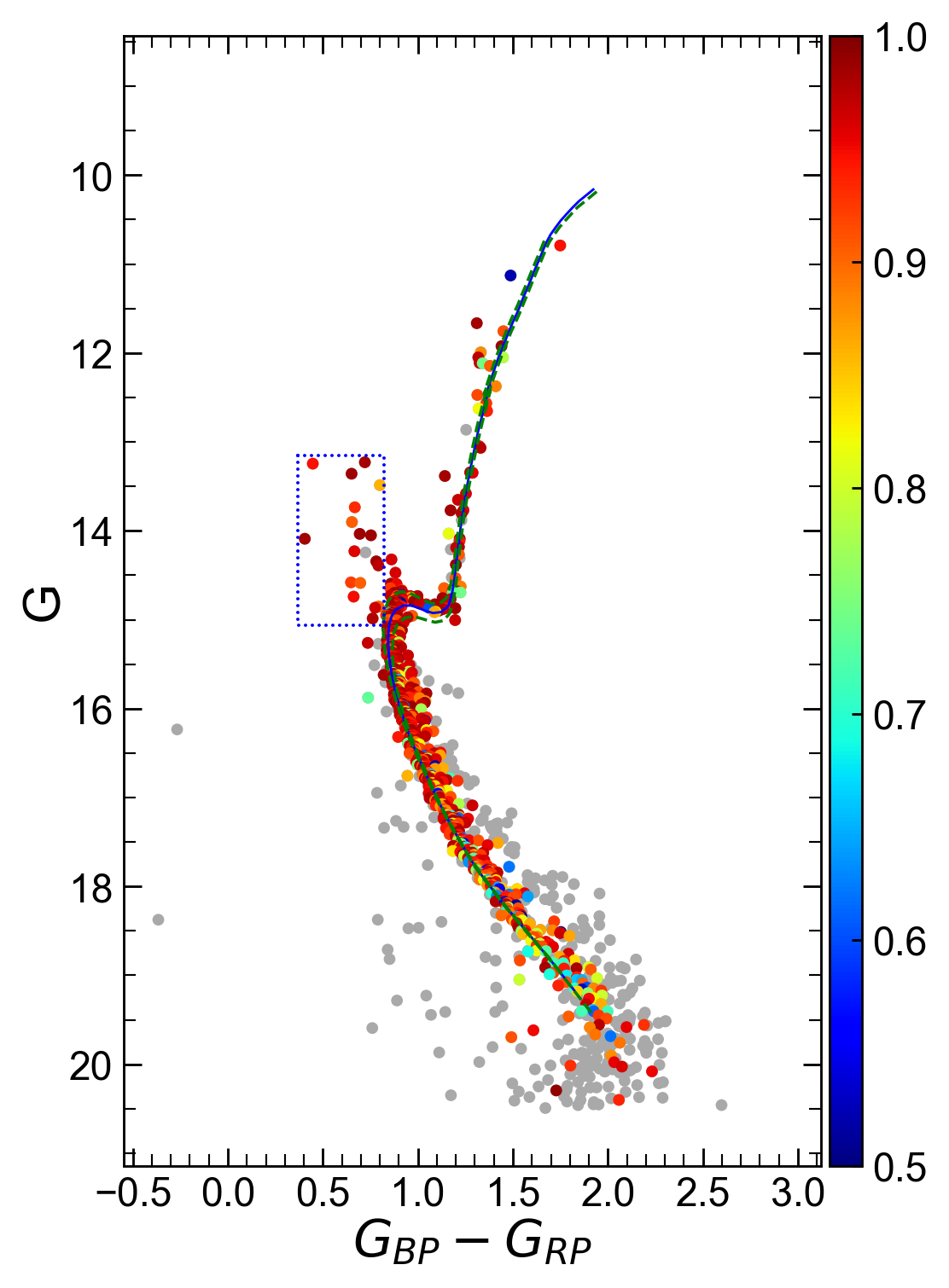}
\caption{CMD of the NGC 188. Different colour scales and colourbar show the membership probabilities of stars with $P\geq 0.5$. Stars with probabilities $P<0.5$ are demonstrated with filled grey circles. The best solution of the fitted isochrones and their errors are inferred as the blue and purple lines, respectively. The age of the blue-lined isochrone matches with 7.65 $\pm$ 1.00 Gyr for the cluster. The BSs were marked within the blue dashed box.}\label{fig:figure_age}
\end{figure}

By keeping metallicity as constant and paying attention to the distance derived from the trigonometric parallaxes we fitted theoretical {\sc PARSEC} isochrones \citep{Bressan_2012} to the CMD and derived age, distance modulus, and reddening simultaneously. Fitting procedure was performed considering the distribution of most likely ($P\geq 0.5$) main-sequence, turn-off and giant members on cluster CMD. The best fitted isochrones of different ages ($t$ = 7.55, 7.65 and 7.75 Gyr) scaled to the mass fraction $z=0.0142$ with distribution of the most likely members on the cluster's $G\times (G_{\rm BP}-G_{\rm RP})$ CMD is shown in Figure \ref{fig:figure_age}. The best fitted isochrones implies the morphology of the cluster in CMD was selected as $t$ = 7.65 $\pm$ 1.00 Gyr. The estimated age is comparable with the values of \citet{Bossini_2019} and \citet{Cantat-Gaudin_2020}.

 The colour excess and isochrone distance values of NGC 188 corresponding to the isochrone age at $z=0.0142$ were obtained as $E(G_{\rm BP}-G_{\rm RP})= 0.066 \pm 0.012$ mag and $d_{\rm iso}= 1806 \pm 21$ pc, respectively. As can be seen from the (see Table~\ref{tab:literature}), colour excess and isochrone distances are consistent with the most of the studies presented by different researchers. The errors in distance modulus and isochrone distance were obtained from the expression of \citet{Carraro_2017}, which takes into consideration the photometric measurements and colour excess with their uncertainties. To carry out more precise comparisons with literature studies, $E(G_{\rm BP}-G_{\rm RP})$ was converted to the $U\!BV$-based colour excess $E(B-V)$ value. For that, we applied the equation of $E(G_{\rm BP}-G_{\rm RP})= 1.41\times E(B-V)$ given by \citet{Sun_2021} and obtained the colour excess as $E(B-V)= 0.047 \pm 0.009$ mag. This result is in good agreement with the values given by \citet{Hunt_2023}, \citet{Cantat-Gaudin_2020}, and \citet{Gao_2022} within the errors (see Table~\ref{tab:literature}). Isochrone distance of NGC 188 derived from isochrone fitting method is agreeable with most studies performed by different researchers (see Table~\ref{tab:literature}) as well as the trigonometric parallax distance, $d_{\rm \varpi}= 1818 \pm 76$ pc, obtained in this study.

\subsubsection{SED Analysis}

\begin{figure}[t!]%Figure 08
\centering\includegraphics[width=1\linewidth]{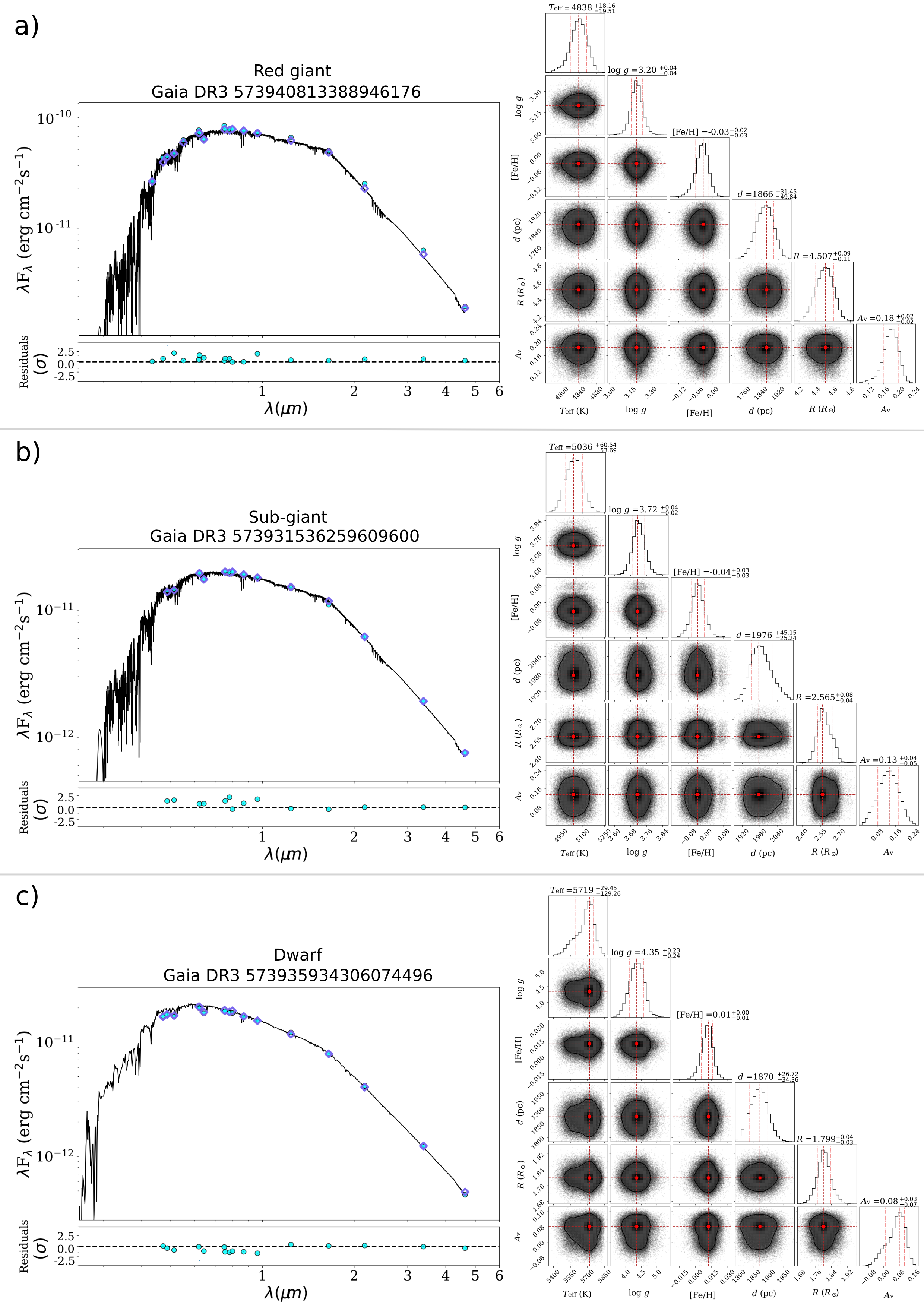}
\caption{SED diagrams (left panels) with best astrophysical parameter solution histograms and distributions (right panels) for three member stars with different luminosity classes. Panels (a), (b) and (c) show the SED analyses for red giant, sub-giant and dwarf member star, respectively.}\label{fig:seds}
\end{figure} 

To conduct Spectral Energy Distribution (SED) analyses for the most likely member stars of NGC 188, flux values measured in various filters across a broad range of the electromagnetic spectrum are essential. As outlined in previous sections, the number of stars with cluster membership $P\geq 0.5$ was established as 868 (see Sec. \ref{sec:Membership}). Since SED analyses focus on determining the basic astrophysical parameters of individual stars, it is necessary to exclude stars in double, multiple, and variable categories among those with high cluster membership from the statistics. To achieve this, the equatorial coordinates and \textit{Gaia} DR3 data of the 868 stars with high cluster membership were considered, and their stellar types and brightness changes were queried through the SIMBAD database.

The query results revealed that 93 stars in the list were classified as double or multiple, 10 were identified as variable stars, and 348 stars lacked sufficient brightness data for SED analysis. Consequently, these stars were excluded from the statistics. The Gaia archive \citep{Gaia_EDR3, Gaia_DR3} includes a Renormalised Unit Weight Error (RUWE) value for each source. This parameter indicates the quality of the astrometric solution for a given source in {\it Gaia}. Ideally, the RUWE value should be around 1.0 for sources where the single-star model fits the astrometric observations well. A value significantly greater than 1.0, such as >1.4, could indicate that the source is non-single or otherwise problematic for the astrometric solution \citep{Fitton_2022}. One giant, one subgiant and three dwarfs were excluded from the analysis after checking the RUWE values of the cluster members. SED analyses were successfully conducted for the remaining 412 single cluster member stars using the \citep[ARIADNE;][]{Vines_2022}, and their basic astrophysical parameters were determined.

The outcomes of the SED analyses for three stars selected from different luminosity classes, along with the cornerplots illustrating the agreement of the main astrophysical parameters, are presented in Figure \ref{fig:seds}. Among the three analyzed stars, the evolved ones exhibited the best fit with the PHOENIX v2 model \citep{Huser_2013}, while the dwarf star demonstrated the best fit with the Castelli and Kurucz model \citep{Castelli_2003}. This agreement is evident from the residual distributions in the bottom panel of the SED distributions for each star. Additionally, the cornerplots in the right panel of the SED distributions for each star indicate the absence of degeneracy between the parameters, with uncertainties at acceptable levels.

\begin{figure}[t!]%Figure 09
\centering\includegraphics[width=1\linewidth]{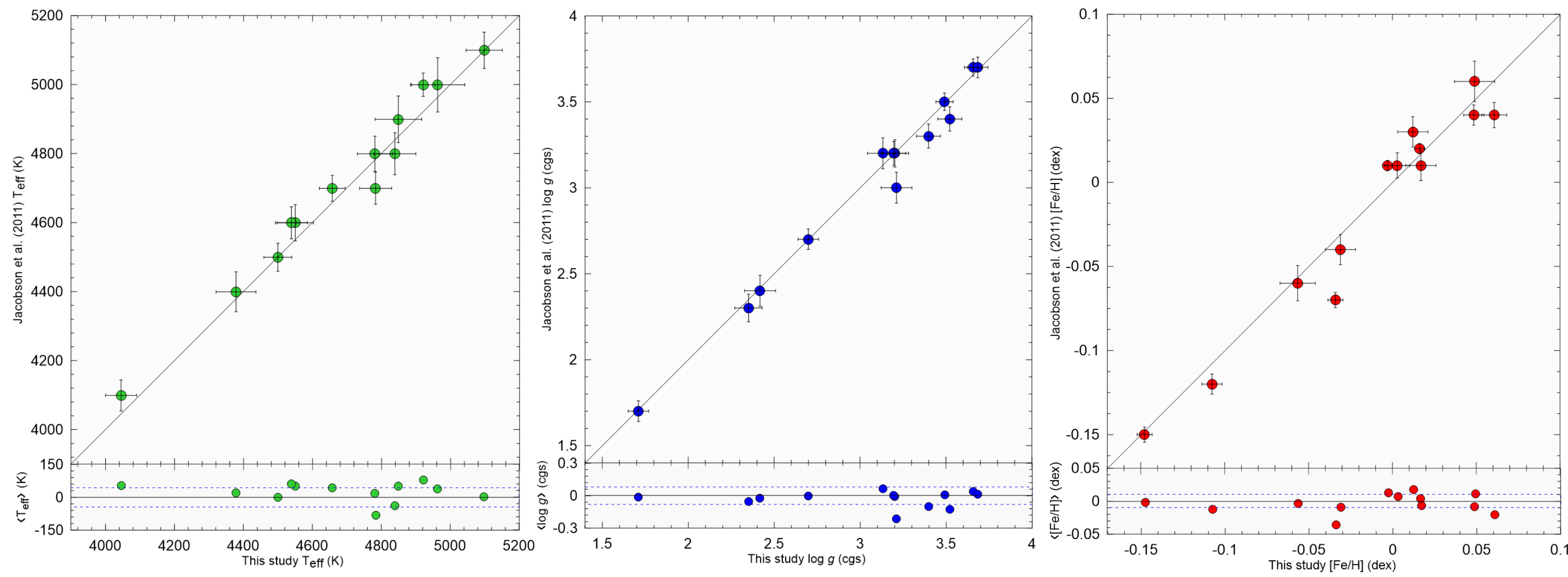}
\caption{The comparison of the astrophysical parameters of the 12 stars spectral analysed by \citet{Jacobson_2011} as members of NGC 188 with the results in this study.}\label{fig:jacobson}
\end{figure} 

To assess the precision of the derived basic astrophysical parameters, we refer to the study by \citep{Jacobson_2011}, who conducted spectral analyses of NGC 188. In their work, \citet{Jacobson_2011} analyzed the chemical abundances of evolved stars in 10 OCs using spectra obtained with the WIYN 3.5m telescope. Examining 31 stars in NGC 188, \citep{Jacobson_2011} identified 12 member stars that are common with the comparison conducted in our study. Among these stars, there are 11 red giants and a subgiant star. The star depicted in the panel of Figure \ref{fig:seds}a was analyzed in both studies. A comparison of the 12 stars, for which model atmosphere parameters ($T_{\rm eff}$, $\log g$, and [Fe/H]) were determined using spectral and SED analysis, is presented in Figure \ref{fig:jacobson}.

In the analyses, the differences in effective temperature, surface gravity, and metal abundance obtained from the two studies, along with the standard deviations of these differences, were calculated as $\langle\Delta T_{\rm eff}\rangle$ = 44 K, $\langle\Delta \log g \rangle$ = 0.08 cgs, and $\langle\Delta$[Fe/H]$\rangle$= 0.01 dex, respectively. The calculated mean differences and standard deviation values being sufficiently small provide crucial evidence that the model atmosphere parameters determined in the two studies are compatible with each other.

With the basic astrophysical parameters of the 412 SED analyzed stars in hand, the absolute magnitudes and reddening-free colour indices of the stars were utilized to ascertain the luminosity classes. This was achieved by creating a more sensitive CMD. The distance relation used to determine the absolute magnitude ($M_{\rm G}$) is given as follows:

\begin{equation}
M_{\rm G} = G_{\rm 0} - 5 \log d + 5,
\end{equation}
where $G_{\rm 0}$ is the de-extinction apparent magnitude of the star and $d$ is the distance determined from the SED analysis. Since SED analysis calculate the extinction value in the V band, selective absorption coefficients ($A_{\rm \lambda}$/$A_{\rm V}$) of 0.83627, 1.08337 and 0.63439 were used for the $G$, $G_{\rm BP}$ and $G_{\rm RP}$ bands, respectively, as defined by the $Gaia$ photometric system \citep{Cardelli_1989}. The following equations were taken into account in the de-extinction of the magnitude:

\begin{eqnarray}
G_{\rm 0} = G - 0.83627 \times A_{\rm V}, \nonumber \\ 
{(G_{\rm BP}})_{\rm 0} = G_{\rm BP} - 1.08337 \times A_{\rm V},\\ \nonumber
{(G_{\rm RP}})_{\rm 0} = G_{\rm RP} - 0.63439 \times A_{\rm V}, \nonumber
\end{eqnarray}

\begin{figure}[t!]%Figure 10
\centering\includegraphics[width=0.6\linewidth]{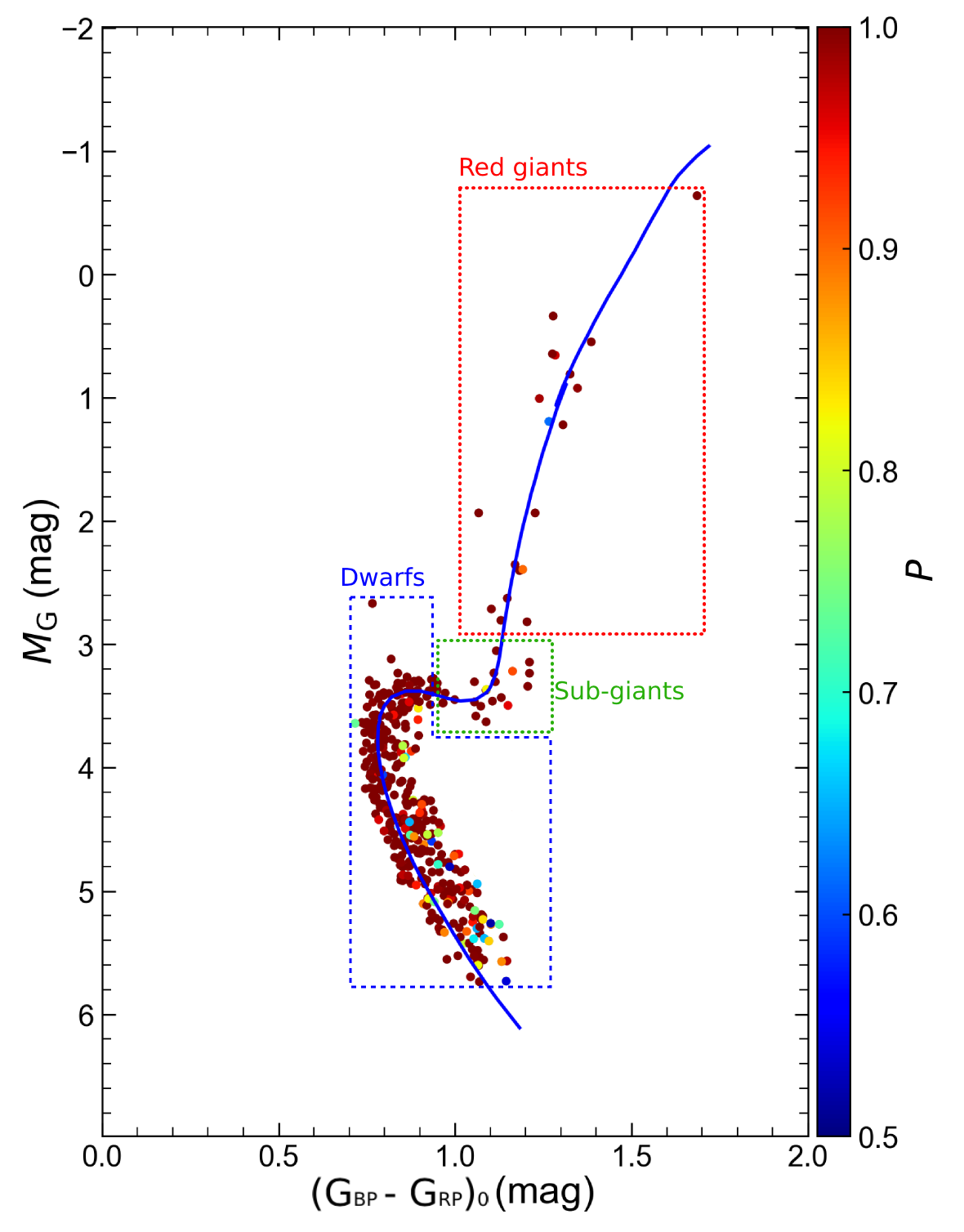}
\caption{The diagram of $M_{\rm G}$ $\times$ ${G_{\rm (BP-RP)}}_0$ for NGC 188. Different colours and colour bar scales indicate the membership probabilities of stars with $P\geq 0.5$. The age of the blue line isochrone matches the age determined by the SED analysis for the cluster 412 members. The red-dashed area denotes red giant stars, the purple dashed area signifies sub-giant stars, and the blue dashed area represents dwarf stars.}\label{fig:lumclass}
\end{figure} 

After calculating the absolute magnitudes and de-reddened colour indices of the cluster member stars, $M_{\rm G} \times (G_{\rm BP}-G_{\rm RP})_0$ diagram was generated Figure \ref{fig:lumclass}. As can be seen from the Figure \ref{fig:lumclass}, the morphology of the cluster is very distinct. The red giant arm of the cluster has an absolute magnitude of -1 < $M_{\rm G}$ (mag) $\leq$ 3 and a colour index ($G_{\rm BP}-G_{\rm RP})_0$ > 0.95 mag, while the lower giant arm has an absolute magnitude of 3 < $M_{\rm G}$ (mag) $\leq$ 3.75 and a colour index of 0.95 < ($G_{\rm BP}-G_{\rm RP})_0$ (mag) $\leq$ 1.30. The remaining stars on the CMD are classified as dwarf stars.

To analyse the differences between the basic astrophysical parameters of the member stars in different luminosity classes of the NGC 188, the ranges of the luminosity classes above were taken into account. The parameter ranges of the stars in each luminosity classes and all member stars analysed by SED are listed in Table \ref{tab:priors}. As can be seen from the bottom row of Table \ref{tab:priors}, the numbers of red giant, sub-giant, and dwarf stars are  20, 18, and 374, respectively.

%Table 03
\begin{table}[h!]
\setlength{\tabcolsep}{10pt}
  \centering
  \caption{Parameters and ranges of values obtained from the best-fit SEDs for 412 member stars of the NGC 188.}
    \begin{tabular}{lcccc}
      \hline
Parameter & Red giants & Sub-giants & Dwarfs & All \\
\hline
    $T_{\rm eff}$ (K) & [4044, 4963]  & [4803, 5507] & [4085, 6075]  & [4044,  6075]  \\
    $\log g$ (cgs)    & [1.50, 3.72]  & [3.40, 3.82] & [3.63, 5.61]  & [1.50,  5.61]  \\
    $\rm [Fe/H]$ (dex)& [-0.11, 0.06] & [-0.07, 0.02]& [-0.09, 0.18] & [-0.11, 0.18]  \\
    $A_{\rm v}$(mag)  & [0.00, 0.25]  & [0.01, 0.18] & [0.00, 0.25]  & [0.00, 0.25]   \\
    $d$ (pc)          & [1766, 1947]  & [1776, 2036] & [1562, 2751]  &  [1562,  2751] \\
    $M$ ($M_{\odot}$) & [0.97, 1.36]  & [0.97, 1.18] & [0.60, 1.24]  &  [0.60,  1.36] \\
    $R$ ($R_{\odot}$) & [3.52, 31.06] & [2.08, 3.28] & [0.82, 2.56]  &  [0.82, 31.06] \\
    $t$ (Gyr)         & [3.26, 10.54] & [5.66, 9.18] & [1.11, 13.39] & [1.11, 13.39]  \\
    \hline   
    $N$               & 20 & 18 & 374 & 412 \\
   \hline
    \end{tabular}%
    \label{tab:priors}%
\end{table}%

When the effective temperature and surface gravity obtained by SED analysis are analysed according to the luminosity classes, it is seen that they are compatible with the stellar evolution models. Evaluating the metal abundance variations across luminosity classes, sub-giant stars exhibit the smallest variation range with $\Delta$[Fe/H] = 0.09 dex, while dwarf stars present the largest variation range with $\Delta$[Fe/H] = 0.28 dex. Analyzing dwarf stars based on their unit absolute luminosity ranges reveals an increase in the range of metal abundances from bright to faint magnitudes. This phenomenon might be attributed to the decreased sensitivity of faint stars in SED analyses.

Considering the $V$-band extinction of the SED-analyzed cluster member stars, a considerable variation is observed, ranging from 0 to 0.25 mag. This significant variation in extinction is evident across all luminosity classes, indicating the presence of differential reddening in the NGC 188 region. Similarly, the distances of SED-analyzed cluster member stars range from 1562 to 2751 pc, with smaller ranges for evolved stars and larger ranges for dwarf stars. This discrepancy is attributed to the relatively decreased probability of cluster membership for fainter dwarf stars, leading to the inclusion of some field stars in the calculations.

Examining the masses of cluster member stars calculated using MESA Isochrones and Stellar Tracks \citep[MIST;][]{Dotter_2016} evolution models as a result of SED analyses, the range is found to be between 0.60 and 1.36 $M_{\odot}$. Evolved stars, as expected, exhibit a range of approximately 0.35 $M_{\odot}$, with the most massive stars falling within this group. Age determinations of the cluster member stars reveal a range between 1.11 and 13.39 Gyr, with dwarf stars exhibiting a large age range consistent with their position in the main-sequence band.

\begin{figure}[t!]%Figure 11
\centering\includegraphics[width=1\linewidth]{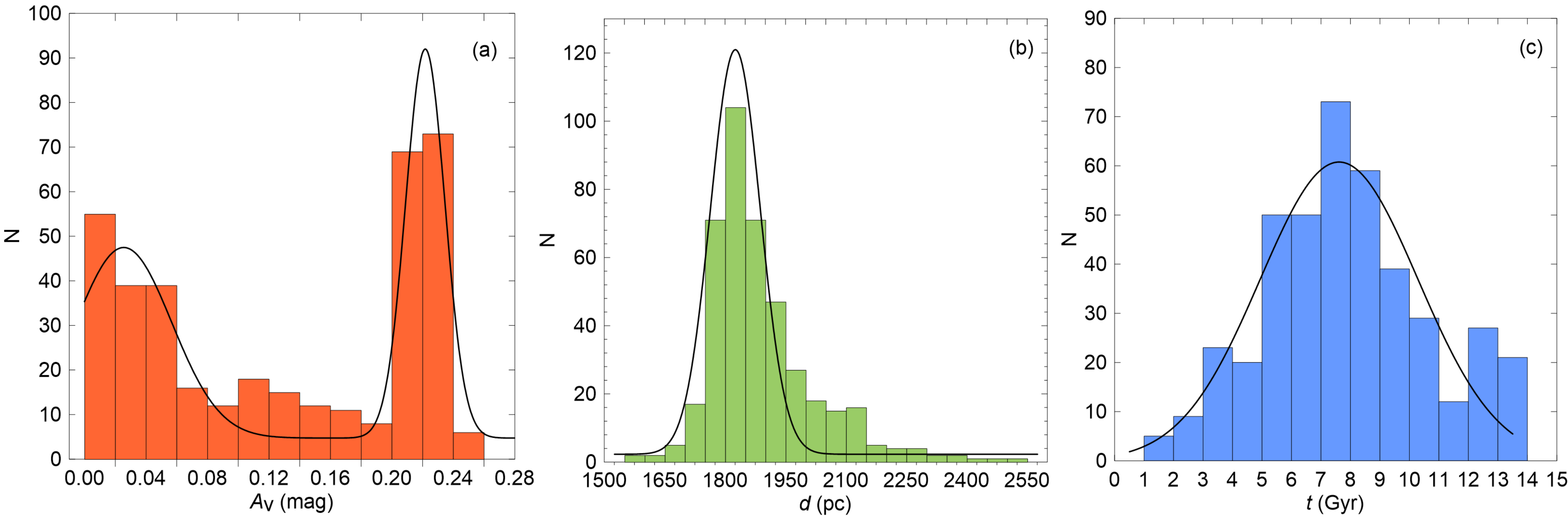}
\caption{Histograms representing the distribution of $A_{\rm v}$, distance ($d$), and age ($t$) values of the 412 members of NGC 188 obtained by SED analysis. Black lines through distributions indicate the standard Gaussian distribution.}\label{fig:histograms}
\end{figure} 

The most uncertain parameters used for age determination are the extinction/colour-excess and distances. The histograms of the $V$-band extinction, distance, and age parameters obtained from the SED analyses of the member stars in NGC 188 are shown in Figure \ref{fig:histograms}a. As seen in Table \ref{tab:priors}, the extinction values of the stars were found to be in a wide range between 0 and 0.25 mag and it was suggested that a differential reddening might be possible. Indeed, when the $V$-band extinction histogram is analysed, a bi-modal distribution is found (Figure \ref{fig:histograms}a). A bi-modal fit was made to this distribution and the mode values were calculated to be $A_{\rm V,1}=0.026\pm 0.025$ mag and $A_{\rm V,2}=0.223\pm 0.017$ mag. This is evidence of differential extinction. When the distance histogram of all stars in the sample is analysed, it shows a distribution that can almost be described by a Gaussian distribution (Figure \ref{fig:histograms}b). When a Gaussian fit is applied to the distribution, the most likely distance of the cluster is found to be 1855 $\pm$ 6 pc. Similarly, the age histogram of all stars is expressed by a Gaussian distribution and the most likely age of the cluster is calculated as 7.61 $\pm$ 0.23 Gyr (Figure \ref{fig:histograms}c).

\begin{figure}[t!]%Figure 12
\centering\includegraphics[width=1\linewidth]{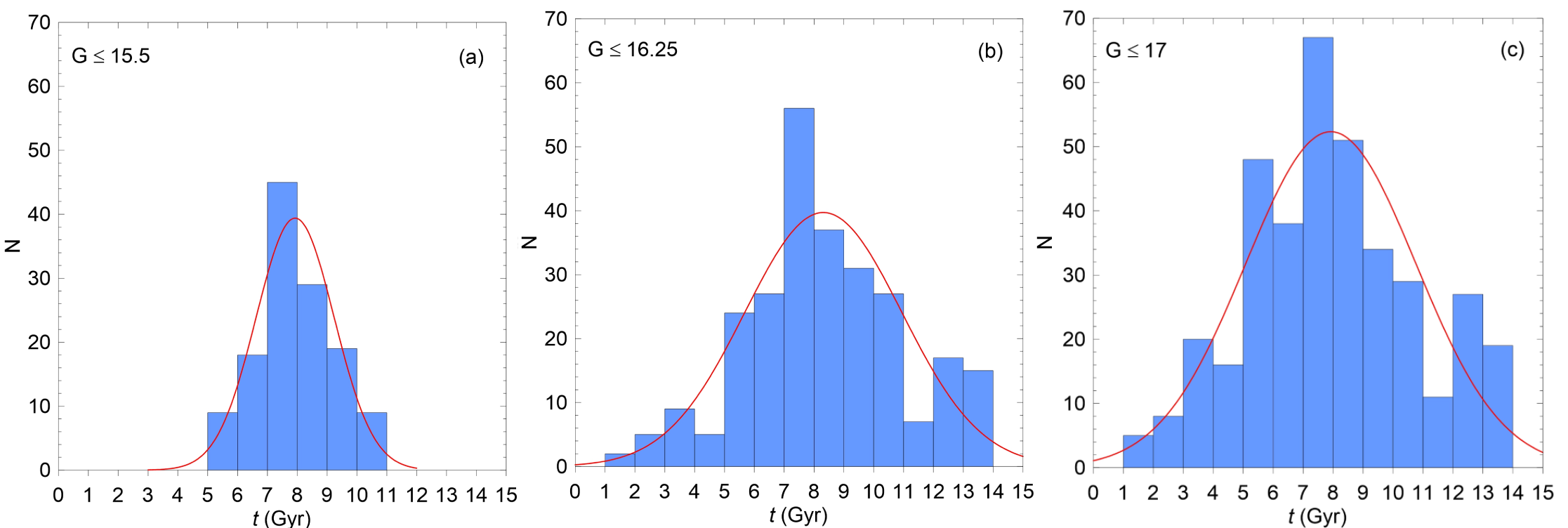}
\caption{Age histograms of main-sequence stars in three different $G$ apparent-magnitude ranges: $G\leq15.5$ (a), $G\leq16.25$ (b), and  $G\leq17$ mag. Red lines through distributions indicate the standard Gaussian distribution.}
\label{fig:histograms_age}
\end{figure}

While the values of the extinction and distance parameters obtained from the SED analyses are concentrated in a narrow range, the range of the calculated ages is quite wide (Figure \ref{fig:histograms}). Since the stars in this study are members of open clusters, their extinction, distances and ages are expected to be in a narrow range. However, the distribution of ages calculated in the analyses is wider than expected. This may be due to the fact that the stars used in the SED analyses are in a wide range of apparent sizes. Taking into account the increase in uncertainties in the results of the SED analyses of faint stars, the ages were recalculated by dividing the sample of stars studied into three different subgroups in the $G\leq15.5$, $G\leq16.25$,  and $G\leq17$ mag interval. The results are given in Table \ref{tab:lumclass} and the age histograms at different apparent magnitude limits are shown in (Figure \ref{fig:histograms_age}). Analysis of the histograms in Figure \ref{fig:histograms_age} shows that the mode values of the ages are very close in all three histograms, but the age distribution widens when faint stars are included in the calculations. This shows that the parameters obtained from SED analysis should be carefully evaluated, especially in open cluster studies, as fainter luminosities are included in the analyses.

Table \ref{tab:lumclass} presents the median values of metal abundances, $V$-band extinction values, distance, and age for stars in different luminosity classes along with their errors. The last row of Table \ref{tab:lumclass} summarizes the median values calculated for 412 stars in NGC 188. The mean metal abundance is $\langle$[Fe/H]$\rangle$ = $0.00 \pm 0.03$ dex, the mean $V$-band extinction value is $\langle A_{\rm V} \rangle = 0.11 \pm 0.09$ mag, the average distance is $\langle d \rangle = 1854\pm 148$ pc, and the mean age is $\langle t \rangle = 7.78 \pm 0.23$ Gyr. Fitting the appropriate {\sc PARSEC} isochrones to the CMD in Figure \ref{fig:lumclass} by considering the values [Fe/H], $A_{\rm V}$, $d$, and $t$ from the last row of Table \ref{tab:lumclass} reveals precise representation of the entire CMD morphology.

Comparing the median distance and age parameters of cluster member stars, it is observed that the results align closely with the values calculated with the Gaussian curve. However, the median $V$-band extinction obtained from SED analysis does not exactly agree due to the presence of differential extinction in NGC 188.

\begin{table}%Table 04
\setlength{\tabcolsep}{10pt}
    \centering
    \caption{Mean values and errors of metallicity ([Fe/H]), $V$-band extinction ($A_{\rm v}$), distance ($d$) and age ($t$) obtained from SED analysis of NGC 188 member stars according to luminosity class.}
    \begin{tabular}{lccccc}
    \hline
        Region & $N$& [Fe/H] & $A_{\rm v}$ & $d$ & $t$ \\ 
               &   & (dex) & (mag) & (pc)  &(Gyr)   \\ 

        \hline
        Red giants & ~~20 &  -0.02$\pm$0.04 & 0.14$\pm$0.06 & 1841$\pm$45  & 6.94$\pm$0.45 \\ 
        Sub-giants & ~~18 &  -0.03$\pm$0.02 & 0.14$\pm$0.05 & 1880$\pm$67  & 6.93$\pm$0.26 \\ 
        \hline
        Dwarfs ($G\leq$15.5) & 129 & ~0.01$\pm$0.03 & 0.10$\pm$0.09 & 1870$\pm$95 & 7.84$\pm$0.13\\
        ~~~~~~~~~~~~ ($G\leq$16.25)&262  & ~0.00$\pm$0.03 & 0.11$\pm$0.09 & 1852$\pm$127 & 8.10$\pm$0.24 \\
       ~~~~~~~~~~~~ ($G\leq$17)& 374  &  0.00$\pm$0.03 & 0.09$\pm$0.10 & 1855$\pm$154 &7.81$\pm$0.30 \\
         \hline
               All & 412   & ~0.00$\pm$0.03 & 0.11$\pm$0.09 & 1854$\pm$148 & 7.78$\pm$0.23 \\
     \hline
    \end{tabular} \label{tab:lumclass}
\end{table}

\begin{figure}[t!]% FIGURE 13
\centering\includegraphics[width=0.8\linewidth]{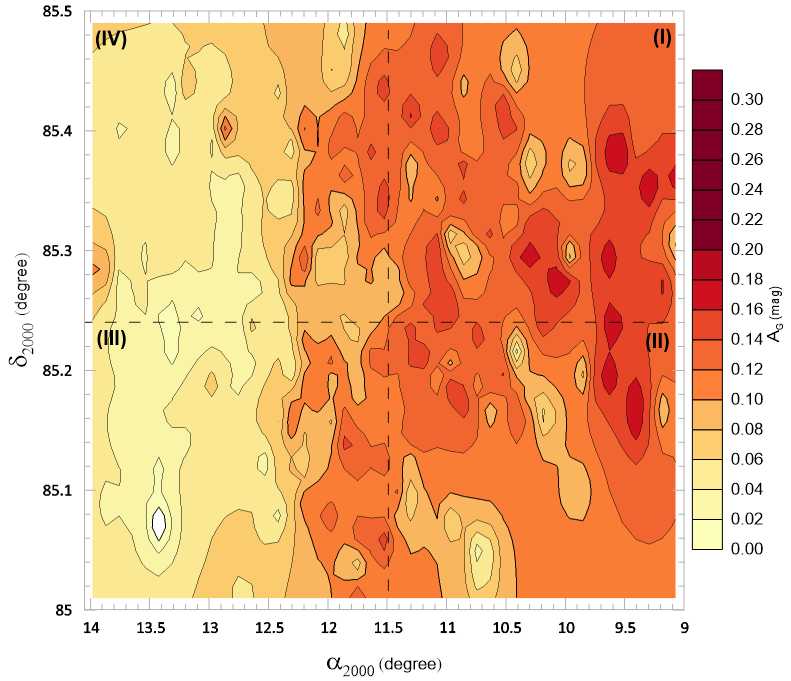}
\caption{The contour plot of the $A_{\rm v}$ values of 412 members of NGC 188 obtained by SED analysis. The center of the cluster is marked by the midpoint of the dashed line.}\label{fig:av}
\end{figure}

To scrutinize the differential extinction within the cluster region more thoroughly, we contoured the $V$-band extinction values calculated from SED analysis for the 412 stars with high membership in the open cluster NGC 188, considering their positions in equatorial coordinates (Figure \ref{fig:av}). Notably, the $V$-band extinctions in the upper right and lower right quadrants of the cluster center exhibit significant differences from those in the upper left and lower left quadrants. The analyses of $V$-band extinction, progressing in a clockwise direction, yield $\langle A_{\rm V} \rangle$ = 0.171 mag in region I, $\langle A_{\rm V} \rangle$ = 0.042 mag in region II, $\langle A_{\rm V} \rangle$ = 0.175 mag in region III, and $\langle A_{\rm V} \rangle$ = 0.045 mag in region IV. This observation underscores that stars with smaller right ascension values in the open cluster NGC 188 tend to have larger $A_{\rm V}$ values.

\subsection{The Blue Straggler Stars}

Blue Straggler Stars (BSSs) found within open clusters deviate from the typical aging trajectory, displaying characteristics that make them appear younger and bluer compared to their counterparts in the surrounding region. Unlike the majority of stars in open clusters that follow established evolutionary pathways, BSSs challenge these norms within the cluster environment. The primary mechanisms contributing to BSS formation involve interactions within binary star systems and stellar collisions occurring in the dense cluster environment \citep{Zinn_1976, Hills_1976}. Theoretical frameworks propose mass gain through stellar collisions, inner binary mergers, or mass transfer during red giant phases, and ongoing research continues to explore these mechanisms \citep{Webbink_1976, Leonard_1989}. In Figure \ref{fig:figure_age}, the blue box highlights 19 stars with cluster membership $P\geq 0.5$ located on the blue side of the cluster's turn-off point, identifying them as high-probability BSSs in NGC 188.

In a study by \citet{Rain_2021}, 22 BSSs were identified using {\it Gaia} DR2 \citep{Gaia_DR2} photometric and astrometric data. Since the membership analyses in this study are based on {\it Gaia} DR3 data, and we considered stars within the limiting radius ($r_{\rm lim}\leq 15\arcmin$), two BSSs identified by \citet{Rain_2021} fall outside these limitations. The BSSs are depicted in Figure \ref{fig:figure_age}. Given that the formation mechanisms of BSSs are primarily associated with mass transfer in close binary systems \citep{McCrea_1964} and stellar collisions \citep{Hills_1976}, we did not include these stars in the SED analysis.

\subsection{Kinematics and Dynamical Orbit Parameters}

In order to determine the Galactic populations of OCs, it is imperative to conduct kinematic and dynamical analyses of their orbits \citep{Tasdemir_2023, Yontan-Canbay_2023}. Detailed kinematic analyses of NGC 188 were carried out, encompassing the determination of its space velocity components, Galactic orbit parameters, and birth radii. These analyses utilized the {\sc MWPotential2014} model from the Galactic dynamics library {\sc galpy}\footnote{See also https://galpy.readthedocs.io/en/v1.5.0/} package by \citet{Bovy_2015}, implemented in the Python programming language. The galactocentric distance and orbital velocity of the Sun were set to $R_{\rm gc}=8$ kpc and $V_{\rm rot}=220$ km s$^{-1}$, respectively \citep{Bovy_2015, Bovy_2012}. The distance of the Sun from the Galactic plane was considered as $Z_0 = 25\pm5$ pc \citep{Juric_2008}. Radial velocity is a crucial parameter for constructing the orbit of a celestial object around the Galactic center. The mean radial velocity of NGC 188 was calculated, taking into account the most likely cluster members with available radial velocity measurements in {\it Gaia} DR3. 68 stars were identified for this calculation. The mean radial velocity was determined using the equation provided by \citet{2022_Carrera}, yielding $V_{\rm R}=$-$41.6 \pm 0.12$ km s$^{-1}$. This result aligns well with findings from literature studies (see also Table~\ref{tab:literature}). To perform the orbit integration of NGC 188, the following parameters were used as input:

\begin{figure}[t!]% FIGURE 13
\centering\includegraphics[width=1\linewidth]{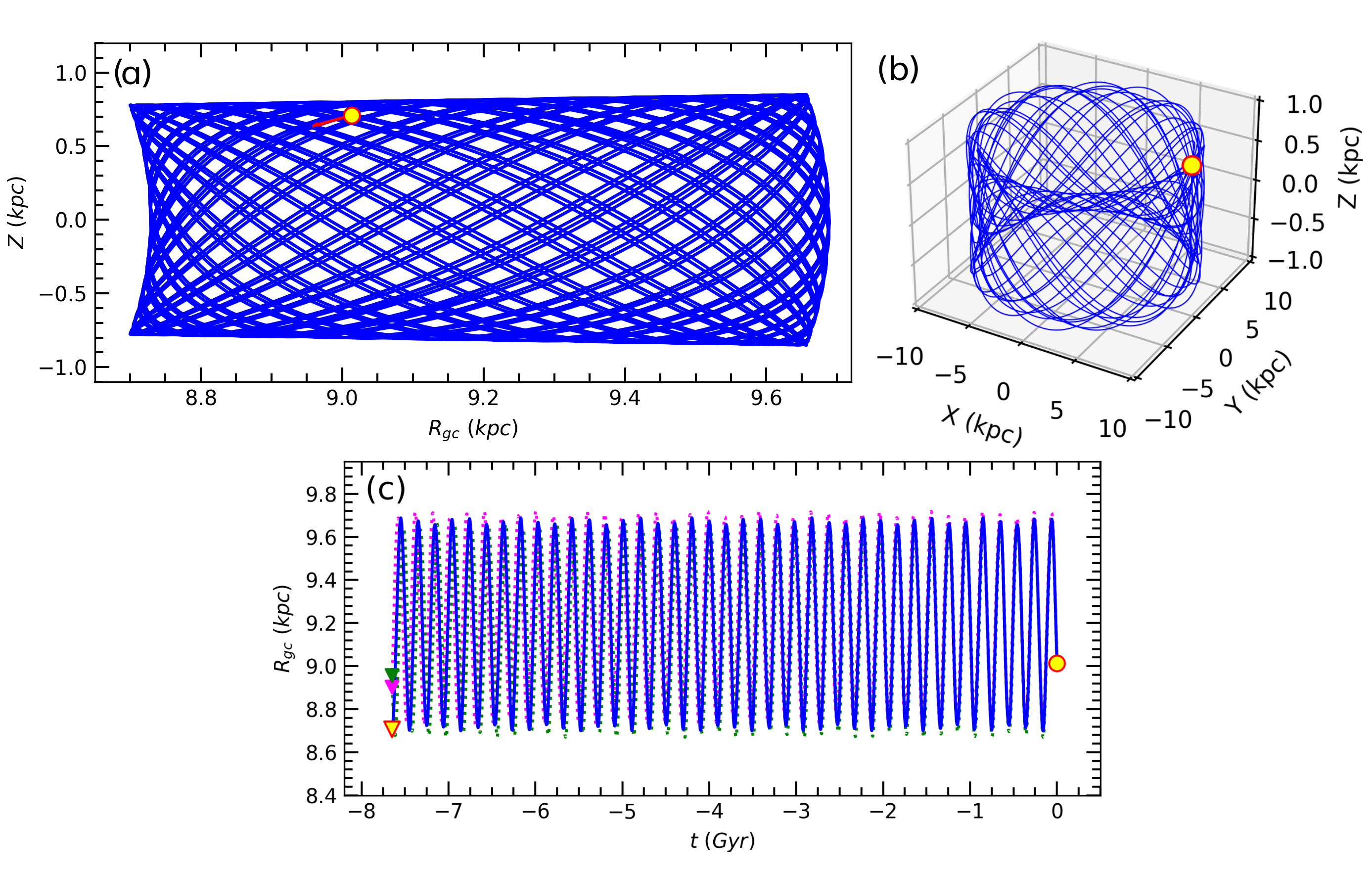}
\caption{The Galactic orbits and birth radii of the NGC 188 are illustrated on three different planes: $Z$ × $R_{\rm gc}$ (a), $X$ × $Y$ × $Z$ (b), and $R_{\rm gc}$ × $t$ (c). Present-day positions are denoted by filled yellow circles, while birth positions are indicated by filled triangles. The red arrow represents the motion vector of the cluster. Additionally, purple and pink dotted lines represent the orbit under consideration of errors in input parameters, with purple and pink-filled triangles that indicating the lower and upper error estimates of the open cluster's birth locations, respectively.}\label{fig:origin}
\end{figure} 

The central equatorial coordinates ($\langle\alpha, \delta\rangle) = (00^{\rm h} 47^{\rm m} 20^{\rm s}.96, \delta= +85^{\circ} 15^{\rm '} 05^{\rm''}.27$) \citep{Hunt_2023}, the mean proper-motion components ($\mu_{\alpha}\cos\delta = $-2.314$ \pm 0.002$, $\mu_{\delta}= $-$1.022 \pm 0.002$ mas yr$^{-1}$) determined in Section 4.2, the isochrone distance ($d_{\rm iso}=1806\pm 21$ pc) from Section 4.3, and the radial velocity ($V_{\rm R}= $-$41.6 \pm 0.12$ km s$^{-1}$) calculated in the study (see also Table~\ref{tab:Final_table}). 
To infer the current likely position of NGC 188, the orbit of the cluster was integrated forward with an integration step from 5 Myr to 7.65 Gyr. Results of orbit integration process: apogalactic ($R_{\rm a}= 9694 \pm 30$ pc) and perigalactic ($R_{\rm p}= 8729 \pm 31$ pc) distances, eccentricity ($e = 0.05$), maximum vertical distance from Galactic plane ($Z_{\rm max}$= 851 pc), space velocity components ($U$, $V$, $W$ = $35.90\pm 0.13, -18.82\pm0.25, $-$23.58\pm0.03$ km s$^{-1}$), and orbital period ($P_{\rm orb}$ = 259 Myr). Taking into account the space velocity component values $(U, V, W)_{\odot}=(8.83 \pm 0.24, 14.19 \pm 0.34, 6.57 \pm 0.21$) km s$^{-1}$ of \citet{Coskunoglu_2011}, we applied a Local Standard of Rest (LSR) correction to the $(U, V, W)$ components of NGC 188. Hence, we derived the LSR corrected space velocity components as $(U, V, W)_{\rm LSR}$ = ($44.73 \pm 0.27, $-$4.63 \pm 0.42, $-17.01$ \pm 0.21$) km s$^{-1}$. Using these LSR results, we estimated the total space velocity as $S_{\rm LSR}$ = $48.08 \pm 0.54$ km s$^{-1}$ (see also Table~\ref{tab:Final_table}). The cluster reaches a maximum distance above the Galactic plane at $Z_{\rm max}$ = $851 \pm 10$ pc, indicating that NGC 188 belongs to the old thin-disc component of the Milky Way \citep{Ak_2015}.

The 3D motion of the cluster around the Galactic center is depicted in Figure \ref{fig:origin}b. As observed from the figure, NGC 188 follows an almost circular orbit around the Galactic plane, experiencing a separation from the Galactic plane by $\pm$ 0.8 kpc during its motion. Figure \ref{fig:origin}c illustrates the distance of the cluster on the $R_{\rm gc} \times t$ plane as a function of time, providing insights into how uncertainties in the input parameters impact the orbit of the cluster. Dynamical analysis reveals that NGC 188 was formed outside the solar circle, with a birth radius of $R_{\rm Birth}= 8.71\pm 0.01$ kpc.

\section{Conclusion}\label{sec:Conclusion}

In this study, detailed analyses of NGC 188 open cluster were performed by using Gaia DR3 astrometric, photometric and spectroscopic data. We identified 868 most likely members for the cluster. Astrophysical parameters were derived via isocron fitting procedure to the CMD. Also we investigated orbit of the NGC 188 by utilizing kinematic and dinamical analyses. In addition, except for similar cluster studies in the literature, the basic astrophysical parameters of 412 most likely members ($P\geq 0.5$) stars brighter than $G$ = 17 mag were determined by SED analyses. The basic astrophysical parameters for NGC 188 were also obtained from the mean values of the SED analysis results and they were compared with those obtained from the isochron-fitting method. We concluded that, the parameters determined from two methods are in good agreement. However, we observed a wide range of metallicity and $V$-band extinction values among the member stars, particularly in NGC 188, where we identified differential extinction for the first time in this study.

All parameters determined in the study are listed in Table~\ref{tab:Final_table}. The main results of the study are summarized as follows:

\begin{enumerate}
\item{From the RDP analyses, we determined the limiting radius by visual inspection as $r_{\rm lim}^{\rm obs}=15^{'}$.} 

\item{Taking into account the results of the photometric completeness limit, the membership probability analyses, and the limiting radius, we identified 868 most likely members with probabilities of $P\geq0.5$ for NGC 188. These stars were used in the cluster analyses.}

\item{The mean proper-motion components were obtained as ($\mu_{\alpha}\cos \delta, \mu_{\delta})=($-2.314$ \pm 0.002, $-1.022 $\pm 0.002$) mas yr$^{-1}$.}

\item{19 most likely BSS members were identified within the limiting radius of the NGC 188.}

\item{The metallicity value for the cluster was taken as $[{\rm Fe/H}]=$ -$0.030 \pm 0.015$ dex, which is presented by \citet{Casamiquela_2021}. We transformed this value into the mass fraction $z=0.0142$ and kept it as a constant parameter for the age and distance estimation.}

\item{The isochrone fitting distance of NGC 188 was determined as $d_{\rm iso}=1806\pm 21$ pc. This value is supported by the distance $d_{\varpi}$= $1818\pm76$ pc that is derived from mean trigonometric parallax. The SED analysis distance of the member of c stars in NGC 188 was obtained as $d=1854\pm 148$ pc.}

\item{The iscohrone fitting method gives the age of the NGC 188 cluster as $t=7.65\pm 1.00$ Gyr while the SED analysis provides the mean age of the cluster determined $t=7.78\pm 0.23$ Gyr}

\item{Orbit integration was performed via {\sc MWPotential2014} model. We concluded that NGC 188 orbits in a boxy pattern outside of the solar circle, as well as the cluster is a member of the thin-disc component of the Milky Way. Moreover, the birth radius ($8.71\pm 0.01$ kpc) shows that the forming region of the cluster is outside the solar circle.}

\item{NGC 188 $V$-band extinction analysis of 412 stars revealed distinct extinction patterns across the cluster's equatorial coordinates. Notably, the upper right and lower right quadrants displayed considerable deviation from the upper left and lower left ones. The examination resulted in varied $V$-band extinction averages, with a clear trend: stars with lower right ascension exhibited higher $A_{\rm V}$ values, indicating a notable correlation within the cluster.}

\item{SED analysis of the member stars revealed that determined effective temperatures and surface gravities align well with stellar evolution models across different luminosity classes.}

\item{The SED analysis revealed an age range of 1.11 to 13.39 Gyr. When dividing the main-sequence stars into three luminosity groups, it becomes apparent that the bright stars exhibit a narrow age range, whereas the range widens as we move towards the faint stars. This implies that the age values of faint open cluster stars require careful evaluation during SED analyses.}

In this study, SED analyses of NGC 188 which is an old open cluster demonstrate that with the increase in the number of photometric data, the fundamental astrophysical parameters of open clusters can be determined with greater precision.

\end{enumerate}

\begin{table}[t!]% TABLE 05
	\renewcommand{\arraystretch}{1.1}
	\setlength{\tabcolsep}{14pt}
	\centering
	\caption{~Fundamental parameters of NGC 188.}
	{\normalsize
		\begin{tabular}{@{}lcc@{}}
			\hline
\multicolumn{1}{c}{Parameter} & Classic Method & SED Method\\
			\hline
            \hline
\multicolumn{1}{c}{Astrometric Parameters}&&\\
            \hline   
			($\alpha,~\delta)_{\rm J2000}$ (Sexagesimal)& \multicolumn{2}{c}{00:47:20.96, $+$85:15:05.27} \\
			($l, b)_{\rm J2000}$ (Decimal)              & \multicolumn{2}{c}{122.8368, $22.3730$}\\    
			$f_{\rm bg}$ (stars arcmin$^{-2}$)                & \multicolumn{2}{c}{$2.832\pm 0.356$}  \\
			$f_{0}$ (stars arcmin$^{-2}$)         & \multicolumn{2}{c}{$12.229\pm 0.768$}    \\
			$r_{\rm c}$ (arcmin)                        & \multicolumn{2}{c}{$2.183\pm 0.304$}    \\
			$r_{\rm lim}$ (arcmin)                      & \multicolumn{2}{c}{15}                 \\
			$r$ (pc)                                    & \multicolumn{2}{c}{7.88}                     \\
			Cluster members ($P\geq0.5$)                & 868                  &  412   \\
			$\mu_{\alpha}\cos \delta$ (mas yr$^{-1}$)   & \multicolumn{2}{c}{-$2.314  \pm 0.002$}    \\
			$\mu_{\delta}$ (mas yr$^{-1}$)              & \multicolumn{2}{c}{-$1.022  \pm 0.002$}   \\
            $\varpi$ (mas)                                  & \multicolumn{2}{c}{$0.550 \pm 0.023$}    \\
			$d_{\varpi}$ (pc)                           & \multicolumn{2}{c}{$1818\pm 76$}          \\
			\hline
\multicolumn{1}{c}{Astrophysical Parameters} & & \\
            \hline
                $E(B-V)$ (mag)                              & $0.047 \pm 0.009$       &  $0.034\pm 0.030$       \\
			$E(G_{\rm BP}-G_{\rm RP})$ (mag)            & $0.066\pm 0.012$        &   ---                   \\
   			$A_{\rm V}$ (mag)                           & $0.146\pm 0.068$        &   $0.107\pm 0.091$      \\
			$A_{\rm G}$ (mag)                           & $0.123\pm 0.022$        &   ---                   \\
			$[{\rm Fe/H}]$ (dex)                        & -$0.030\pm 0.015$$^{*}$ &   $0.00\pm0.03$         \\
			Age (Gyr)                                   & $7.65\pm 1.00$          &  $7.78\pm 0.23$         \\
			$V-M_{\rm V}$ (mag)                         & ---                     &  $11.306  \pm 0.007$    \\
            $G-M_{\rm G}$ (mag)                             & $11.407  \pm 0.025$     &  ----                   \\
			$d_{\rm iso}$ (pc)                          & $1806 \pm 21$           &  $1854\pm 148$          \\
                $(X, Y, Z)_{\odot}$ (pc)                    & (-$906$, $1403$, $687$) & (-$915$, $1418$, $695$) \\
			$R_{\rm gc}$ (pc)                           & 9015                    & 9027                    \\
            \hline
\multicolumn{1}{c}{Kinematic \& Dynamic Orbit Parameters} & & \\
            \hline
                $V_{\rm R}$ (km s$^{-1}$)                   & \multicolumn{2}{c}{-$41.60 \pm 0.12$} \\
			$U_{\rm LSR}$ (km s$^{-1}$)                 & \multicolumn{2}{c}{$+44.73 \pm 0.27$} \\
			$V_{\rm LSR}$ (kms$^{-1}$)                  & \multicolumn{2}{c}{-$4.63 \pm 0.42$}  \\
			$W_{\rm LSR}$ (kms$^{-1}$)                  & \multicolumn{2}{c}{-$17.01 \pm0.21$}  \\
			$S_{_{\rm LSR}}$ (kms$^{-1}$)               & \multicolumn{2}{c}{$48.08 \pm0.54$}   \\
			$R_{\rm a}$ (pc)                            & \multicolumn{2}{c}{$9694\pm 30$}      \\
			$R_{\rm p}$ (pc)                            & \multicolumn{2}{c}{$8729 \pm 31$}     \\
			$z_{\rm max}$ (pc)                          & \multicolumn{2}{c}{$851\pm 10$}       \\
			$e$                                         & \multicolumn{2}{c}{$0.052\pm 0.001$}  \\
			$P_{\rm orb}$ (Myr)                         & \multicolumn{2}{c}{$259\pm 10$}       \\
			$R_{\rm Birth}$ (kpc)                        & \multicolumn{2}{c}{$8.71 \pm 0.01$}  \\
			\hline
			$^{*}$\citet{Casamiquela_2021}
		\end{tabular}%
	} %ends the small font selection
	\label{tab:Final_table}%
\end{table}%

\vspace{8pt}

\section*{Acknowledgements}
We are grateful to the referees for their valuable feedback, 
which improved the paper. This study has been supported in part by the Scientific and Technological Research Council (T\"UB\.ITAK) 122F109. Special thanks to Dr. Olcay Plevne for technical support. This research has made use of the WEBDA database, operated at the Department of Theoretical Physics and Astrophysics of the Masaryk University. We also made use of NASA's Astrophysics Data System as well as the VizieR and Simbad databases at CDS, Strasbourg, France and data from the European Space Agency (ESA) mission \emph{Gaia}\footnote{https://www.cosmos.esa.int/gaia}, processed by the \emph{Gaia} Data Processing and Analysis Consortium (DPAC)\footnote{https://www.cosmos.esa.int/web/gaia/dpac/consortium}. Funding for DPAC has been provided by national institutions, in particular, the institutions participating in the \emph{Gaia} Multilateral Agreement.


%\bibliographystyle{mnras}

\bibliographystyle{mnras}
\bibliography{refs}

% Don't change these lines
\bsp	% typesetting comment
\label{lastpage}
\end{document}